\documentclass{jfm}
\usepackage{graphicx}
\usepackage{newtxtext}
\usepackage{newtxmath}
\usepackage{natbib}
\usepackage{hyperref}
\hypersetup{
    colorlinks = true,
    urlcolor   = blue,
    citecolor  = black,
}

\newcommand{\RomanNumeralCaps}[1]
\linenumbers

\title{Convection Velocity in Turbulent Boundary Layers under Adverse Pressure Gradient}

\author{Artur Dróżdż\aff{1}
  \corresp{\email{artur.drozdz@pcz.pl}},
  Paweł Niegodajew\aff{1}
  Mathias Romańczyk\aff{1}
 \and Witold Elsner\aff{1}} 

\affiliation{\aff{1} Department of Thermal Machinery, Czestochowa University of Technology, Armii Krajowej 21, 42-200 Czestochowa, Poland}

\begin{document}
\maketitle

\begin{abstract}

The convection velocity ($U_C$) of turbulent structures has been studied in adverse-pressure-gradient (APG) turbulent boundary layers (TBLs) for a wide range of Reynolds numbers $Re_\tau \approx1400-4000$. The study is based on estimation of the convection velocity using decomposed streamwise skewness factor introduced in (Dróżdż A., Elsner W., Int. J. of Heat and Fluid Flow. 63 (2017) 67–74) and verified by means of two-point correlation method. It was shown that in the overlapping region of APG flows, the convection velocity profiles (when scaled in viscous units) reassemble the universal logarithmic law characteristic for the ZPG flows up to Clauser-Rotta pressure gradient parameter $\beta \lesssim 19$ for the considered range of Reynolds number, what means that in the inner region of TBL the friction velocity in APG is not proportional to $U$ (as in ZPG) but to $U_C$ instead. The physical mechanism that explains the impact of increased convection velocity on the mean flow is proposed. The difference between $U_C$ and $U$ increases as a function of APG, which causes the stronger sweeping that enhances momentum transfer to the wall and compensates the weaker mean shear profile that is created by lower vorticity near the wall. This effect is a result of an enhancement of amplitude modulation of the small scales by large scale motion. The process becomes more pronounced as eddy density grows, so with increasing Re. The proposed model addresses a number of literature observations found in adverse pressure gradient flows which have been so far left without a well-founded explanation. 

\end{abstract}



\section{Introduction}
\label{sec:intro}

A comprehensive knowledge about the streamwise convection velocity $U_C$ of vortical structures is essential to obtain a deep understanding of physics of adverse-pressure-gradient (APG) turbulent boundary layer (TBL) flows. The spanwise vortex, also known as the hairpin head, according to the model of \cite{Adrian2000} (featuring hairpin, cane and arch vortices), is the main element of a complex tree-dimensional vortical structure of the TBL inner region \citep{Robinson1991}. The contribution of this vortex is important as the vorticity associated with it adds to produce the mean shear profile in TBL \citep{Elsinga2012}. According to \cite{Kim2006}, hairpin heads are accompanied with turbulent quadrant  sweep and ejection events also known as burst-like events i.e. the negative Reynolds shear stress $(-\overline{uv})$. Sudden lift up and ejection of near-wall fluid are associated with second-quadrant while sweep of higher momentum fluid towards the wall is associated with fourth quadrant Reynolds stress generation \citep{Blackwelder1983}. As pointed out by \cite{Schroder2011} these events may propagate together with a vortex. A couple of vortices are usually combined into packets that create a shear layer between the uniform momentum zones above and below the packet \citep{Laskari2018}. One should note that there are some confusing inconsistencies in nomenclature regarding ‘sweep’ and ‘ejection’ terms, since they have been also applied to events associated with large-scale high- and low- momentum zones  located further away from the wall \citep{Hutchins2012a}. The structures’ organisation model proposed by \cite{Stanislas2017} clearly indicates the difference between these events. 

\cite{Drozdz2017Uc,Drozdz2017tp} found that in APG flow the convection velocity of small-scale pairs of sweep and ejection events is 40\% higher than the local mean velocity $U$ in the buffer region. This observation was further confirmed by \cite{Balantrapu2021}, who based on comparison of the two-point correlations from PIV with the single-point hot-wire estimates, using Taylor’s hypothesis \citep{Taylor1938} showed a similar increase in $U_C$ in the inner part of TBL. The incresed $Uc$ may lead to imbalance between sweep and ejection small-scale events which affects the wall-normal momentum transfer.

The characteristic convection velocity of the turbulence is not easy to determine due to following reasons: (i) different parts of a structure can travel with different convection velocities, (ii) as a structure evolves, its overall transport velocity may not be the same within its lifetime, (iii) structures with different sizes propagate with different velocities, and (iv) $U_C$ also depends on the actual wall-normal location of a structure within TBL \citep{Krogstad1998}. Most of available research works were focused on the convection velocity in canonical turbulent wall-bounded flows \citep{Krogstad1998,Osterlund2003,DelAlamo2009,Buxton2011a,Elsinga2012,Schlatter2012,DeKat2013,Atkinson2014,Renard2015,Yang2018,Liu2020}, where two-point correlations was used to estimate $U_C$ for the turbulent coherent vortices. An example of such studies may be the work of \cite{Krogstad1998}, where the authors using two-point correlation method showed that the mean $U_C$ is almost the same as mean velocity $U$. \cite{Krogstad1998} showed also that the distance between the hot-wire probes can potentially affect the estimation of $U_C$. Namely, if this distance is too short, the resultant convection velocity can be underestimated and can also vary due to the prongs wake effect generated by the upstream probe. This effect together with a spatial averaging of the probe \citep{Marusic2010a} are the main reasons for an inconsistency between the $U_C$ results obtained in earlier works \citep{Moin2009}.
Detailed information concerning the scale-dependent (or wavenumber- $(k_x)$ or frequency-dependent) $U_C$ was provided by \cite{DelAlamo2009} based on direct numerical simulation (DNS) of channel flows and later, by \cite{Renard2015}, based on large eddy simulations (LES) of the TBL developing in zero pressure gradient (ZPG) at high Reynolds numbers \Rey. They found that the structures with the streamwise wavelength, $\lambda_x < 2\delta$ (where $\delta$ is the boundary layer thickness) propagate with the velocity close to the mean (averaged across-scales) $U_C$. On the other hand, $U_C$ of the large-scale motion (LSM), $\lambda_x > 2\delta$, was found to vary relatively little with wall-normal direction $y$. In another work \cite{Renard2015} showed that very-large-scale motions (VLSM) or superstructures \citep{Kim1999a} \citep{Hutchins2007c}), with $\lambda_x > 10\delta$, convect at uniform velocity within the whole TBL (please note that further in the text LSM referes to all scales with $\lambda_x > \delta$). The large-scale motions are extended to higher distance from the wall and, therefore, propagate faster than small-scales in the near-wall region. 
On the other hand, the small-scales tend to propagate with the local mean velocity except near the wall, below $y^+\approx 10$ where $U_C^+\approx 10$ \citep{DelAlamo2009,Krogstad1998,Renard2015} (where $^+$ refers to viscous units). It is because, similarly as for LSMs, their cores are located at higher wall-normal viscous units $y^+\approx 10$. \cite{Renard2015} demonstrated that the mean $U_C$ (instead of varying across-scales $U_C$) in the inner layer obtained from the two-point correlation can be used to estimate the spatial premultiplied energy spectra. In the outer layer of the channel flows, the streamwise convection velocity is always slightly lower than $U$ because structures convect also in the wall-normal direction \citep{Kim1999a}. The research of \cite{Osterlund2003} suggests that $U_C^+$ near the wall increases with an increasing $Re$. However, more recent paper by \cite{Liu2020}, indicates that $U_C^+$ is invariant of $Re$. Based on the above review, one can assume that for the canonical flows the mean convection velocity of the small scales can be approximated using the mean flow velocity in the logarithmic region. This assumption is also valid for large scales, but only in the location where the core of the LSM crosses the log layer in its geometrical centre \citep{Chung2010}.

In adverse pressure gradient (APG) flows, which are of considerable practical interest, \cite{Drozdz2017Uc,Drozdz2017tp,Drozdz2021a,Balantrapu2021} found that the $U_C$ of the small scales is higher with respect to the mean flow up to the upper bound of the inner region, while the highest difference was reported in the buffer layer (reaching values that are even twice as high as the mean velocity at strong pressure gradients). 
Increased $U_C$ in the inner region is most probably induced by the LSM as it forces the production of the small-scale turbulence in the high-speed low-momentum zones in the process of the amplitude modulation \citep{Mathis2009}. In this process, the amplitude of the small-scale signal is modulated by LSMs that accords to the quasisteady quasihomogenous theory of \cite{Zhang2016}. \cite{Agostini2016} found that the positive (sweeps) and negative (ejection) envelopes of the small-scale motions are asymmetrically modulated by the LSMs. The process occurs due to variation of the friction velocity induced by the large scales, especially for high $Re$ flows and/or in the APG \citep{Harun2013}. \cite{Ganapathisubramani2012} showed that not only the amplitude is modulated, but also the occurrence of the small scales. 
Change of the occurrence was assigned to the effect of the frequency modulation of small scales or/and to the change in $U_C$ that occurs due to the high- and low-speed LSMs. Both effects arise with a certain time lag with respect to the LSM \citep{Ganapathisubramani2012}. Later on, \cite{Baars2015}, using a wavelet analysis, confirmed that in the inner region the change of occurrence  is accompanied with frequency modulation. In another works \citep{Baars2017a,Tanarro2020}, the authors found that the $U_C$ variation is consistent with the quasisteady quasihomogenous theory of \cite{Zhang2016}. Additionally, \cite{Lee2014} and \cite{Lozano-Duran2014} showed that the local convection velocity depends on the sign and magnitude of the instantaneous local value of large-scale velocity fluctuation. 

Summing up, one may conclude that large scales modulate the convection velocity; however, the mean flow remains unchanged according to the theory of \cite{Zhang2016}. This hypothesis is not fully consistent as the envelopes of small-scale amplitude and frequency modulation lead the large-scale flow pattern in the inner region of TBLs (see Figure 12 in the paper of  \cite{Baars2015}). Later, it was found that the shifts of those envelopes in respect to LSM is even more pronounced \citep{Iacobello2021} when utilising time-varying $U_C$ instead of averaged across-scales $U_C$ \citep{Yang2018}. Therefore, it may indicate that small-scale turbulence that convects faster decelerates when falling into the low-speed zone from denser populated high-speed zone. Finally, the increased mean convection may enhance the wall-normal momentum transfer, especially in the low-momentum zones as sweep events are stronger in that region (phase lag effect). 

The present work concerns the comprehensive experimental study of the $U_C$ in the near-wall region of the TBL under strong APG conditions, which has not been previously studied in detail. Convection velocities were estimated using the method proposed by \cite{Drozdz2017tp} that utilises the cross-product term of the decomposed skewness factor introduced by \cite{Schlatter2010} and \cite{Mathis2011}. Validity of this method has been confirmed using the two-point correlation measurements in the vicinity of separation. Since the validity of the method relying on skewness factor was demonstrated, $U_C$ was determined for three different experimental databases of \cite{Drozdz2021a} \cite{Drozdz2017Re} and \cite{Monty2011}. These databases contains well-resolved single-point hot-wire velocity signals for a wide range of pressure gradients and different flow histories. Detailed analysis of the convection velocity across the boundary layer thickness for different pressure conditions allowed to formulate the generalised conceptual model of the impact of the $U_C$ on the quadrant events and consequently on the mean flow in APG.

\section{Methodology }

Although the present paper is focused mainly on estimation of mean $U_C$ using indirect method based on decomposition of skewness factor of streamwise fluctuation from available databases, the new experiment was preformed concerning two-point correlation measurements which gives direct estimation of $U_C$ and verification of indirect method. 

\subsection{Experimental test section}
\begin{figure}
  \centerline{\includegraphics[width=0.9\textwidth]{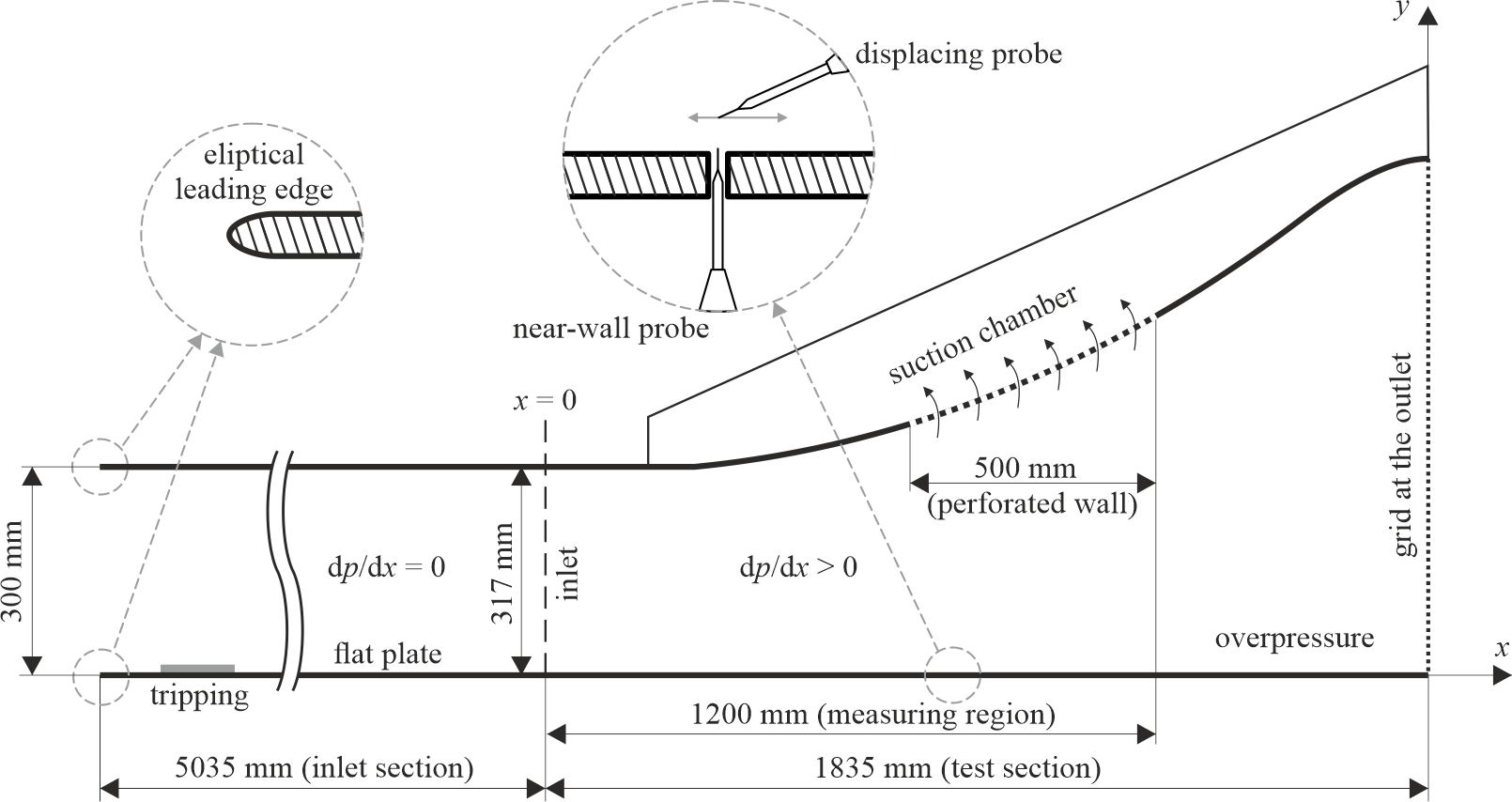}}
  \caption{Test section geometry}
\label{fig:stand}
\end{figure}

The experimental study was conducted in an open-circuit wind tunnel with APG test section schematically shown in Figure \ref{fig:stand}. The high Reynolds number ZPG TBL was development on a 5035 mm long flat-plate in the inlet section, where the friction Reynolds number $Re_{\tau} = 1400 - 4000$. The free-stream turbulence intensity at the inlet plate ($x = 0$ in Figure \ref{fig:stand}), just before the APG test section, was at the level of 0.7\%. 

To obtain a TBL close to the separation at the bottom flat plate, the suction of the flow at the distance of 500 mm in the streamwise direction was applied through the perforated upper wall (with 10$\%$ perforation by surface area - 0.5 mm holes) and its exact location is presented in Figure \ref{fig:stand}. A more detailed description of the experimental stand can be found in our previous work \citep{Drozdz2021a}.

A hot-wire anemometry (HWA) Streamline-Pro DANTEC system was used. Two identical hot-wire probes were used (modified 55P31 – 0.4 mm in length and 3 \textmu m in diameter) with the sampling frequency from 10 kHz for $Re_{\tau} = 1400$ to 20 kHz for $Re_{\tau} = 4000$. The free-stream velocity, static pressure at the inlet plane $(x = 0)$, and pressure at the suction chamber were monitored with 1$\%$, 10$\%$, and 2.5$\%$ uncertainty levels, respectively. 
The friction velocity $u_\tau$ was calculated using the corrected Clauser-chart method (CCCM) (proposed by \cite{Niegodajew2019}) for APG flows up to the shape factor $H = \frac{\delta^*}{\theta} < 2.0$ (where $\delta^*$ is the displacement thickness and $\theta$  the momentum-loss thickness). Above $H=2.0$ another correlation was used relaying on $H$, also according to \cite{Niegodajew2019}. The uncertainty of the friction velocity estimation was up to ~2.5\% for profiles characterised by $H < 2.0$ and up to ${\sim} 5\%$ for $H \geq 2.0$. Uncertainties in the measurements/calculations of $U$ and $H$ were below the level of 1.0\% and 1.5\%, respectively. Uncertainty of the hot-wire position was  $\Delta y = \pm 0.02$ mm and  $\Delta x = \pm 0.0375$ mm in the wall-normal direction and in the streamwise direction, respectively.

\subsection{Methodology of direct convection velocity estimation}
\label{CV-measure}

The two-point correlation based convection velocity estimation was used in the present work to verify the approach that utilises decomposition of the skewness factor proposed by \cite{Drozdz2017Uc}. To avoid the wake effect behind the upstream probe that was previously observed by Krogstad et al. (1998) and the probes’ interference during the measurements, which is especially important for flows close to the separation, one of the probes was plugged into the wall (submerged in the sublayer). It assures that the near-wall probe readings resemble the wall shear stress signature \citep{Baars2017a}. The second (displacing) probe was positioned above the first one. The location of each probe can be seen in the zoomed part of the flat bottom wall in APG section in Figure \ref{fig:stand}). For that specific probes' arrangement, the convection velocity of the near-wall footprint of the vortex, which core passes through the displacing probe location, is possible for self-similar vortex which is coherent with the wall (e.g. wall-attached) \citep{Baars2017a}. 

The output voltages from the two hot-wires were sampled simultaneously and the signal from the displacing probe was post-processed using a wavelet transform: 

\begin{equation}
WT(a_W,b) = \frac{1}{\sqrt{a_W}}\int_{-\infty}^{\infty}  u(t) \phi
\left( \frac{t-b}{a_W} \right) dt
\label{eqW}
\end{equation}

where $a_W$ is the scale and $b$ is the translation of the wavelet function $\phi$. The first derivative of the Gaussian distribution was used as the wavelet function $\phi$. Such an function assures that an opposite pairs of quadrant events are detected. These events correspond to accelerations (abrupt increase in $u(t)$ signal, $+$) and decelerations (abrupt decrease in $u(t)$ signal, $-$) for the selected scale of the wavelet $a_W$. The detected acceleration and deceleration events represents retrograde and prograde vortices, respectively which will be discussed in more details further in this section. In the most cases the prograde vortical structure is the hairpin head while the retrograde vortex can be produced in the inclined shear layer between two consecutive hairpins. 

\begin{figure}
  \centerline{\includegraphics[width=1.0\textwidth]{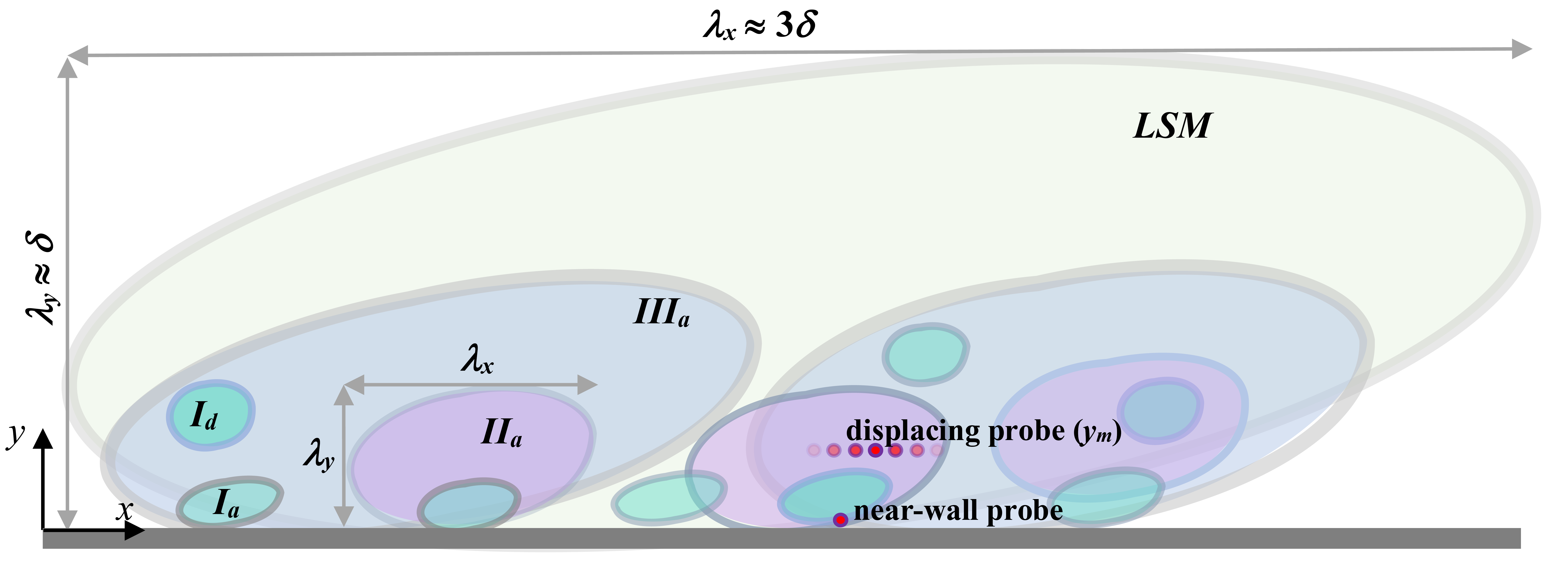}}
  \caption{View of the LSM and the three hierarchies of small-scale eddies in the APG with corresponding locations of the probe}
\label{fig:fig2}
\end{figure}
 
The estimation of $U_C$ for that specific probes arrangement requires the proper selection of the streamwise to wall-normal wavelengths aspect ratio $\lambda_x/\lambda_y$ \citep{Chung2010,Baars2017a} is needed to specify the time scale $\tau_t = \frac{\lambda_x}{U_C}$. In practice, $\tau_t$ is proportional to the distance of the displacing probe from the wall $y_m$. For instance, as it was shown by \cite{Baars2017a}, $\lambda_x/\lambda_y$ based on the linear coherence spectrum is as high as 7 for high $Re$ ZPG flows. However, for analysed APG flow the aspect ratio from linear coherence spectrum is at the level of 4 for all analysed cases in Table \ref{tab:2}. So, the reduction in aspect ratio is barely twice and is closer to that for LSM shown by \cite{Harun2013} and \cite{SanmiguelVila2020} for weak APG for which $\lambda_x/\lambda_y\approx 5$, which is substantially lower than for ZPG, where $y^+=3.9\sqrt{\delta^+}$ and $\lambda_x\approx 6\delta$ that gives the ratio $\lambda_x/\lambda_y\approx50$. However, for the short vortex packet-type structures, which are typical for canonical flows, the $\lambda_x/\lambda_y \simeq 1.5-3$ \citep{DelAlamo2006,Hwang2014}. 
On the other hand \cite{Drozdz2017Re} and \cite{Drozdz2021a} showed that for strong APG $\lambda_x\approx 2.2\delta$ and the ratio equals $\lambda_x/\lambda_y\approx 3.7$.

To understand the process of scale selection, Figure \ref{fig:fig2} shows the concept of the different hierarchies ($I$, $II$, $III$) of wall-attached (subscript $a$) and wall-detached (subscript $d$) eddies. Note that only the small-scale structures can be considered as detached eddies. However, the wall-attached eddies are the main energy-containing structures in the logarithmic and outer regions \citep{Yoon2019}. The sizes of eddies in the logarithmic region are proportional to the distance of their centres from the wall (i.e. $y_c$). The procedure should ensure that only the wall-attached vortical structures are detected (namely, the one marked as $II_a$ type in Figure \ref{fig:fig2}), which have cores at location $y_m$ and are coherent with the wall. It should be noted that for type $I$ eddies’ cores at the location of $y_m$ for which $\lambda_y/2 \ll y_m$ and $y_m \gg y_c$, the coherence with the wall is marginal because it does not touch the wall. In the case of the type $III$ eddy, where $\lambda_y/2 \gg y_m$ and $y_m \ll y_c$, there is a possibility to obtain a coherence with the wall, but its level is low.

\begin{figure}
  \centerline{\includegraphics[width=1.0\textwidth]{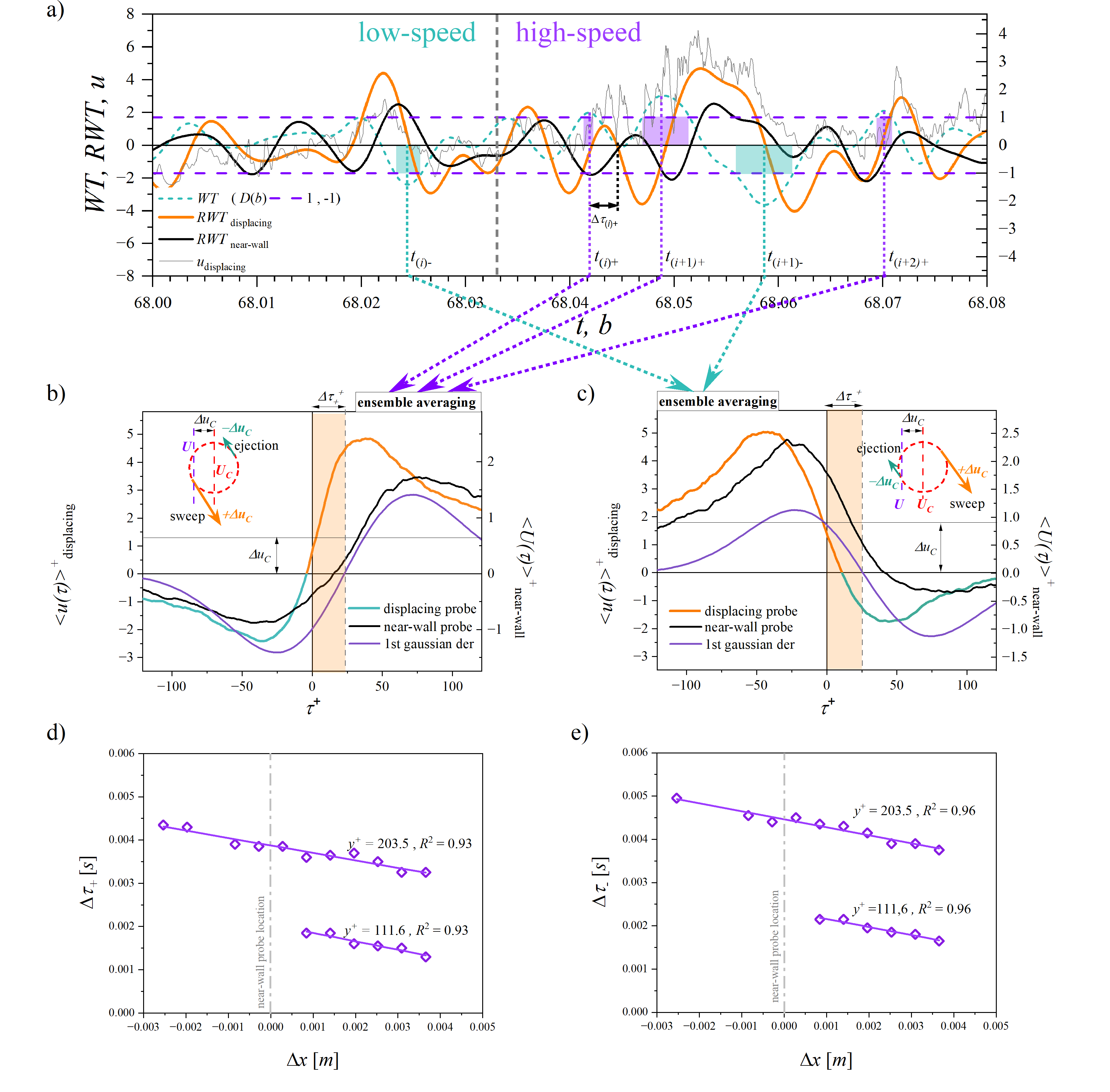}}
  \caption{Detection process: wavelet transform $WT(a_W,b)$ (eq. \ref{eqW}) with corresponding reverse transform $RWT(a_W,b)$ for both signals (a); example of time shift $\Delta\tau_+$ and $\Delta\tau_-$ using eq. \ref{eqW2} in phase-averaged waveforms (eq. \ref{eqp-a}) for accelerations (b) and decelerations (c) ($y^+=111.6$, $\beta = 30.7$, high-$Re$ case). Example of the fitting slopes for accelerations $\Delta(\Delta\tau_+)/\Delta x$ (d) and decelerations $\Delta(\Delta\tau_-)/\Delta x$ (e) ($\beta = 30.7$, high-$Re$ case). Schematic vortex for both detections shows the effect of increased convection ($\Delta u_C$) on sweep and ejection events (color of event corresponds to the color on the displacing probe waveforms)}
\label{fig:fig3} 
\end{figure}

In the present study various $\lambda_x/\lambda_y$ ratios were tested in order to detect the wall-attached small-scale ($\lambda_x<\delta$) eddies only. It was found that $\lambda_x/\lambda_y \simeq 4$ was the most suitable ratio (the least scattering of points in Figures \ref{fig:fig3}d and \ref{fig:fig3}e was achieved). Finally, in the detection process of wall-attached eddies, the scale of the wavelet $a_W=\lambda_x/U_C/4=2y_m/U_C$ was chosen. The detected scale is the small one as $\lambda_x \ll \delta$ (see Table \ref{tab:3}), for which the $U_C$ is very close to the averaged across-scales convection velocity as was pointed out by \citep{DelAlamo2009}. It is clear that the individual structure has also different convection speed, which depends on the interaction with other structures in the surroundings. However, as the present work is focused on analysis of the average across-scale convection velocity of all types of events: ejections, sweeps and vortical structures, it becomes reasonable to choose the mean propagation velocity of vortical structures as its approximation. For type $II$ eddy, the signal waveform corresponds to the one for which a pair of opposite events exists (see Figure \ref{fig:fig3}b) and \ref{fig:fig3}c)). In the case where the core of the vortex omits the location of the displacing probe (type $I$ and $III$ attached eddies), the signal waveform corresponds to a single event only (passage of the edge of the vortex). This event, however, is not included in the present analysis. To be sure that pairs of opposite events are detected (detection of the vortex core that passes through the single-wire probe), the first derivative of the Gaussian function was employed in the wavelet transformation. There is also a possibility to use the second derivative of the Gaussian function, however, it is not suitable to detect single vortical structure as it detects a single sweep or a single ejection event.

It should be noted here that there is a confusing interpretation concerning $U_C$ of different turbulent events. \cite{Krogstad1995} and \cite{Lozano-Duran2014} showed different $U_C$ of sweep, ejection, acceleration and declaration events, while other researchers (i.e., \cite{Schroder2011}) suggested that sweep and ejection propagate with the vortex. However, these two observations are consistent if one considers the vortex stretching. When the spanwise vortex grows larger during their lifetime, the events produced at the front of the vortex moves faster, while at the tail end moves slower. Please note that the retrograde vortices in comparison to prograde exhibit shorter lifetime as they have to face an increasingly strong wall-normal velocity gradient and so, they are progressively destroyed \citep{Herpin2013}. Therefore, only increasing size of prograde vortical structures affects the average $U_C$ of sweep and ejection, which is higher and lower, respectively relative to $U_C$ of the vortex.

The detection process was shown in Figure \ref{fig:fig3}. The displacing probe signal $u_{displacing}$, its wavelet transform $WT(a_W, b)$ and reversed transform $RWT(a_W, b)$ for both two-point correlation signals are shown in Figure \ref{fig:fig3}a. The local extremum of the transform was used as the detection time, while the maximum or minimum of the wavelet transform determines type of event ($+$ or $-$ for detection function shown in Figure \ref{fig:fig3}a as $D(b)=1$ or $D(b)=-1$, respectively). Acceleration ($+$) and deceleration ($-$) events were phase-averaged in both signals (from the near-wall and displacing probes) separately. However, they are aligned to the phase-averaging time of the event detected in the displacing signal, using the following formula:

\begin{equation}
\langle u(\tau)_{\pm} \rangle = \frac{1}{N_{\pm}} \sum_{i}^{N_{\pm}} u_{\pm}(t_i + \tau),
\label{eqp-a}
\end{equation}
%

\noindent where $t$ is the time for $i$-th detection, while $\tau$ is the phase-average time. The sample results of the ensemble averaging are shown in Figures \ref{fig:fig3}b and \ref{fig:fig3}c for acceleration ($+$) and deceleration ($-$) events, respectively. The selected threshold applied on the wavelet transform (see Table \ref{tab:3}) enabled obtaining a high number of detections ($N > 3000$) from the signal (Figure \ref{fig:fig3}a), which was sufficient to get smooth near-wall phase-averaged waveforms (see example in Figures \ref{fig:fig3}b and \ref{fig:fig3}c). The model of the spanwise vortex (see the upper corners of Figuress \ref{fig:fig3}b and \ref{fig:fig3}c) indicates an increased sweep event due to a difference between the convection and mean velocity $(\Delta u_C = U_C-U)$ as previously shown by \cite{Drozdz2014c}. The phase-averaged waveforms for the displacing probe indicate a pair of opposite asymmetrical events and their positive skewness confirm an increased $U_C$ with respect to the mean flow velocity (see also \cite{Drozdz2017Uc}). For both detections, the strong sweep event occurs, which was also observed in many other studies devoted to TBLs under APG conditions \citep{Gungor2021,Mottaghian2018,Nagano1998a,Krogstad1995}.

To estimate the value of $\Delta \tau$ needed for estimation of $U_C = \frac{\Delta x}{\Delta \tau}$, the wavelet transform was also applied on the phase-averaged waveform of the near-wall probe using the relation:

\begin{equation}
WT(2a_W,b)_{\pm} = \frac{1}{\sqrt{2a_W}}\int_{-\infty}^{\infty}  \langle u(\tau)_{\pm} \rangle \phi
\left( \frac{\tau-b}{2a_W} \right) d\tau.
\label{eqW2}
\end{equation}

As the coherent waveforms obtained from near-wall probe are wider than those from displacing probe (see Figures \ref{fig:fig3}b and \ref{fig:fig3}c), the scale of the wavelet applied was set to be equaled to $2a_W$. This is because a slightly different shift in time occurs between displacing probe waveform and the near-wall waveform for each detection ($\Delta\tau$ obtained for a single event), which can be observed in Figure \ref{fig:fig3}a. The phase time $\tau$ for which the square of wavelet transform reaches the maximum value, determines the shift $\Delta \tau$ needed to calculate $U_C$. Having, several streamwise locations of displacing probe separated by $\Delta x$, $U_C$ is calculated based on the slope of $\Delta x = f(\Delta \tau)$ distribution at given $y_m$ (see Figures \ref{fig:fig3}d and \ref{fig:fig3}e). Due to different number of $N_+$ and $N_-$ detections, $U_C$ was calculated as a weighted average of $U_C$ estimated separately for acceleration and deceleration events.

As it can be observed in Figure \ref{fig:fig4}d and \ref{fig:fig4}e, there is a strong coherence between the two-point correlation waveforms, which gives only up to 5\% uncertainty. However, $U_C$ obtained with two-point correlations closest to the wall for the first few upstream positions of the displacing probe (i.e.negative $\Delta x$ at position $y^+ = 111.6$ in Figurs \ref{fig:fig3}d and \ref{fig:fig3}e) are affected due to interactions between probes (as they were too close to each other) and they are not included in the analysis, however, only half range of the streamwise positions was analysed, so the uncertainty level increased up to 10\%.

\subsection{Estimation of mean convection velocity based on the velocity skewness}
\label{Uskew}
As presented in the previous section the estimation of the mean convection velocity is a challenging task as it requires the utilisation of a spatial correlation measurement technique. An alternative approach was proposed by \cite{Drozdz2017Uc}, which employs the streamwise skewness cross-product term $\overline{3\overline{u_S^2u_L}}=\frac{3\overline{u_S^2u_L}}{\overline{u^3}}$ of the decomposed skewness factor, where subscripts $L$ and $S$ denote the large- and small-scale components of the streamwise velocity fluctuations, respectively. An advantage of this approach is that it utilises single-point measurements and the only requirement to estimate $U_C$ is the proper selection of the time cut-off scale to separate large (subscript $L$) and small (subscript $S$) signal components. The wavelength of the cut-off was set as a constant at the local value of $\delta^+$. As spatial to temporal conversion of the cut-off scale employs $U_C$ in the Taylor hypothesis, a correction is needed for the inner layer, where the mean convection velocity differs from the mean velocity especially for APG flows. Therefore, the estimation of the cut-off time scale was performed in two stages. Assuming that the convection velocity in viscous units is constant in the buffer layer and equals $U_C^+ \approx 10$ \citep{DelAlamo2009,Krogstad1998,Renard2015}, the cut-off time scale must be constant as well. Having an estimated value of the cut-off time scale, a preliminary value of the skewness cross-product term $\overline{3\overline{u_S^2u_L}}$ could be calculated. Next, using the relationship:

\begin{equation}
U_C^+ = U^+ + \overline{{3\overline{u_S^2u_L}}} C^+,
\label{eq2}
\end{equation}
%

\noindent which was first introduced by \cite{Drozdz2017Uc}, where $C^+ = 16.34$, the first approximation of $U_C$ can be obtain. In the second stage, the calculation of  $\overline{3\overline{u_S^2u_L}}$ utilises the cut-off time scale $\tau_t^+ = \frac{\lambda_x^+}{U_C^+}$ using $U_C^+$ estimated in the first stage. Then, again using the relation (\ref{eq2}), the final $U_C^+$  was estimated. 
\begin{table}
  \begin{center}
\begin{tabular}{lccccc}
  source        & symbol    &	$\Rey_\tau$            & $\beta$       &	$H $          &	comments\\[3pt]
   \hline
\cite{Drozdz2021a}  &    $ \square$     &	1400        &	7.6 – 120.8 & 1.57 – 2.72   & strong APG\\
                    &	  $\Diamond$    &   4000        & 5.6 – 40.2	& 1.41 – 2.38	& strong APG\\
                 \hline
\cite{Drozdz2017Re} &    $\circ$        &	2800        &	4.8 – 25.2  & 1.41 – 2.11   & moderate APG\\
                \hline
\cite{Monty2011}     &   $\triangle	$   &1800 – 3900    & $\sim4.5$	    & 1.53 – 1.61   & weak APG\\
 \hline
\end{tabular}
  \caption{Time signals database used in the present study}
  \label{tab:1}
  \end{center}
\end{table}

In the outer layer, where $U_C$ is close to the mean velocity, the cross-product term $\overline{3\overline{u_S^2u_L}}$ cannot be used as it becomes strongly negative, the approximation of $U_C$ by mean velocity was used. The constant $C^+=16.34$ was estimated by \cite{Drozdz2017Uc} by fitting the ZPG $U_C$ profiles (estimated by eq.\ref{eq2}) with the data of mean convection velocity estimated by \cite{Krogstad1995}. It is worth mentioning that profiles of $U_C$ \citep{Krogstad1995} agrees well with those calculated by \cite{DelAlamo2009}, which in turn are obtained at the same Reynolds number as for data analysed in \cite{Drozdz2017Uc}. The constant $C^+=16.34$ was then verified for moderate $\beta $ values by \cite{Drozdz2017tp}. 

The sensitivity analysis showed that, when the cut-off wavelength is doubled, the estimated $U_C$ changes by up to 1\% and 10\% for the moderate and strong pressure gradient, respectively. The method proves its excellent performance over a wide range of pressure gradient conditions, however, the verification in the flow close to the separation is necessary, which will be showed in Section \ref{sec:verify}.

\section{Experimental databases}

Although APG TBL is of high practical interest, there is a very few reliable databases that include velocity signals obtained experimentally for flows with a relatively strong pressure gradient. The analysis in this paper is restricted to TBLs at relatively strong velocity deficit and high Reynolds number obtained in three different experiments, (\cite{Drozdz2021a}, \cite {Drozdz2017Re} and \cite{Monty2011}). Other databases in APG flows, which were delivered by \cite{Volino2020}, \cite{Romero2022} and \cite{SanmiguelVila2020a} have not been taken into account due to two reasons. In the first work, the highest Reynolds number was equaled $Re_\tau = 1130$, for which the logarithmic region of the mean velocity profile is hard to observe \citep{Niegodajew2019}, whereas for the reaming cases $\beta$ value were too low, so the difference between $U_C$ and $U$ profiles is barely visible. There is also database of \cite{Balantrapu2021}, however, it was not included in the  analysis as the $u_\tau$ was not determined with a sufficient accuracy. Information about the key flow parameters investigated in the present investigation, including the friction Reynolds number $Re_{\tau} = \frac{u_\tau \delta}{\nu}$, the Clauser–Rota pressure parameter $\beta = \frac{\delta^*}{\tau_w} \frac{dP_e}{dx}$, and $H$ are collected in Table \ref{tab:1}. In these relationships, $U_e$ is the free stream velocity, $\nu$ is the kinematic viscosity, and $P_e$ is the free stream static pressure. It is important to note that, in each considered experiments, $\beta$ increases downstream, so these flow cases should be regarded as non-equilibrium boundary layers. As can be seen from Table \ref{tab:1}, the data covers a wide range of flow parameters, i.e. $Re_\tau$, at the inlet to APG section, ranges from 1400 to 4000, $Re_\theta$ is from 4900 to 23600, $\beta$ is from 4.5 to 120.8, and $H$ is from 1.36 to 2.2. One can also note that databases of \cite{Monty2011}, \cite {Drozdz2017Re} and \cite{Drozdz2021a} can be classified as high Reynolds number flows at strong, moderate and weak APG based on $\beta$ ranges, respectively.

\section{Verification of the convection velocity estimation method based on the skewness decomposition}
\label{sec:verify}
To evaluate the accuracy of the method proposed by \cite{Drozdz2017Uc} the convection velocity estimation approach that relies on two-point correlation supplemented by a wavelet analysis was used (see Section \ref{CV-measure}). The convection velocity from two-point correlation was estimated for profiles close to the separation (in Table \ref{tab:2}) as the method based on decomposed skewness (\ref{eq2}) is most sensitive to chosen cut-off scale $\delta^+$ for that profiles (see Table \ref{tab:2}). Note that the $\delta$, displacement distance ($\Delta x$), maximum displacement distance ($\Delta x_{max}$), and sampling duration ($\Delta t_0^+$) are shown in viscous units ($^+$). This method was used to verify estimation of $U_C$ by eq. \ref{eq2} for single-point data of \cite{Drozdz2021a},

Table \ref{tab:3} presents the complete set of parameters used in the estimation of $U_C$ using the two-point correlations. The parameters in Table \ref{tab:3}, not mentioned earlier, are the momentum thickness based Reynolds number $Re_{\theta} = \frac{u_e \theta}{\nu}$, the threshold of the detection with respect to the streamwise velocity variance $\overline{uu}$, the number of detections $N$ and the ratio of negative $N_-$ to positive $N_+$ number of detections. Please note that the value of $\lambda_x$ is comparable to the integral length scales estimated by \cite{Balantrapu2021}.

\begin{table}
  \begin{center}
\def~{\hphantom{0}}
  \begin{tabular}{lcccccccc}
$\Rey_\theta$    & $\beta$   &	$H$    &  $u_\tau$   &  $\delta^+$  & $\Delta x^+$    & $\Delta x^+_{max}$  & $\Delta t_0$ & comments\\[3pt]
       -        &    -      &     -     &   [m/s]	 &  -      &       -       &            -      &     - & \\
        \hline
 8 000	        &   48.6    &   2.20	&   0.082	 &  805    &   2.98	     &   41.7	        & 0.047	& low-\Rey\\
  \hline
 26 341	        &   23.6	&   1.82    &	0.430	 &  3 900  &  15.5	        & 217	            & 0.217 &	high-\Rey\\
 \hline
 29 048	        &   30.7	&   2.04	&   0.360	 &  3 660  &  13.0	         & 195	            & 0.166	&	high-\Rey\\
 \hline
  \end{tabular}
  \caption{Parameters of two-point correlation measurements }
  \label{tab:2}
  \end{center}
\end{table}

\begin{figure}
  \centerline{\includegraphics[width=0.6\textwidth]{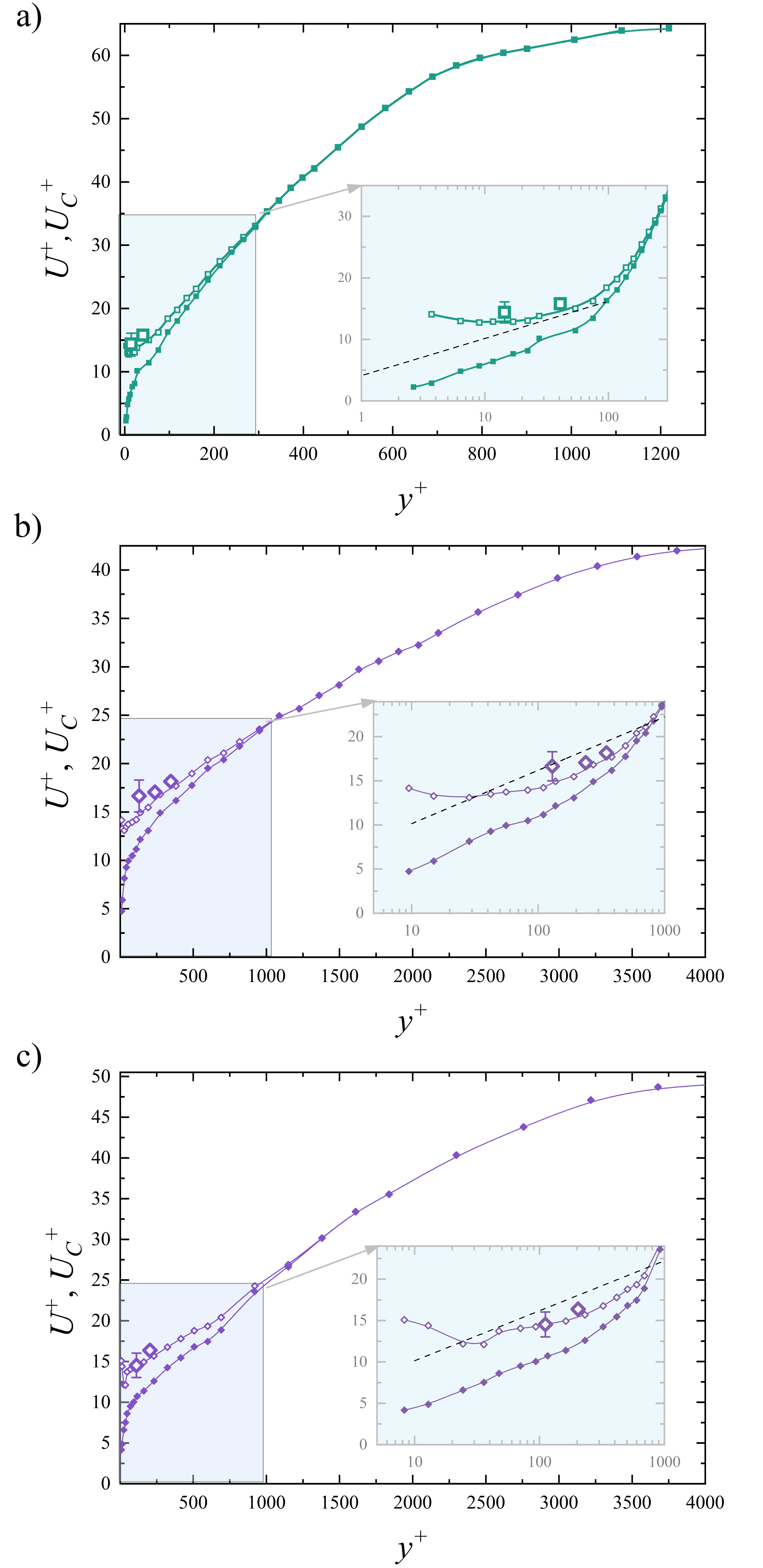}}
  \caption{Verification of the estimated convection velocity: (a) low-$Re$; (b and c) high-$Re$. Small open symbols represent the estimation using Eq.\ref{eq2} and the large open symbols depicts the estimation with two-point correlations; solid symbols represent the mean streamwise velocity. Symbols are as in Table 2 (where the size of the large symbols denote the uncertainty level for the full range distance $\Delta x_{max}^+$, and additional error bars are introduced for the nearest points to the wall that indicate an increased uncertainty, since half of the $\Delta x_{max}^+$ range was used to avoid probes’ interference). Dashed line represents the logarithmic line, in which $\kappa = 0.38$ and $B = 4.1 $}
\label{fig:fig4}
\end{figure}

The values of $U_C$, which are based on two alternative approaches, are presented together with $U$ profile in Figure \ref{fig:fig4}a for low-$Re$ and in Figures \ref{fig:fig4}b and \ref{fig:fig4}c for high-$Re$ conditions. As can be seen, $U_C$ (open symbols) close to the wall is substantially higher than $U$ (solid symbols). Small open symbols represent an estimation of $U_C$ using eq. \ref{eq2} and large open symbols depicts results from the two-point correlations. The inner region is presented in semi-log scale in the zoomed-in figure, where the dashed line represents the logarithmic line:

\begin{equation}
U^+ = \frac{1}{\kappa} ln(y^+) + B,
\label{eq3}
\end{equation}
%

\noindent where the von Karman $\kappa = 0.38$ and $B = 4.1$ are universal constants are valid for a wide range of $Re$ \citep{Nagib2008}. The uncertainty level of  $U_C^+$ from the two-point correlations was estimated to be not higher than 5\% except for the first points closest to the wall for the high-$Re$ case, where the uncertainty level equals up to 10\% (marked with bars in Figure \ref{fig:fig4}). This value also includes the uncertainty that results from the step motor positioning in the streamwise direction, which was 2.5\%. Please note that for the profiles presented in Table \ref{tab:2} $\delta^+$ values are much lower than at the beginning of APG (see $Re_\tau$ in Table \ref{tab:1}). It is important to note that the local value of $\delta^+$ should be used to calculate the cross product term in eq. \ref{eq2}, otherwise $U_C$ cloud be overestimated close to separation.

The most relevant information that can be attained from Figure \ref{fig:fig4} is that the estimation of $U_C^+$, using the single-point measurement with the decomposition of the skewness factor, accurately reflects the mean convection velocity distribution of the small scales estimated from the two-point correlations. Note that the identical value of $C^+=16.34$ was used as in the previous work \citep{Drozdz2017Uc,Drozdz2017tp}. Therefore, it can be stated that the above method is correct even for high pressure gradient flows, where a non-zero value of the mean wall-normal component of velocity is likely to occur \citep{Vinuesa2018}.
The convection velocity distributions shown in Figure \ref{fig:fig4} also confirm the previous findings of \cite{Drozdz2017Uc,Drozdz2017tp} and \cite{Balantrapu2021} concerning increased $U_C$ by 40\% or even higher with respect to the mean flow in the buffer layer.

\begin{table}
  \begin{center}
\def~{\hphantom{0}}
  \begin{tabular}{lcccccccccc}
$\Rey_\theta$    &  $\beta$   &	$y_m^+$     &   $\lambda_{y}^+$ & $\lambda_{x}^+$   & $\tau_t^+$    & $a_W^+$ & threshold &	$N$	   & $N_-/N_+$  & comments\\[3pt]
  \hline
8 000  & 48.6 &  14	&	28	&	112	&	9	&	2.2	&	$0.15\overline{uu}$	&	4344	&	0.93 &
 low-\Rey\\
8 000  & 48.6 &	41	&	82	&	328	&	22	&	5.5	&	$0.15\overline{uu}$	&	3386	&	0.99 & low-\Rey\\
\hline
26 341 & 23.6 & 129	&	258	& 1032	&	68	&	17	&	$0.3 \overline{uu}$	&	5078	&	0.66 & high-\Rey\\
26 341 & 23.6 &	240	&	480	& 1920	&	118	&	29	&	$0.3 \overline{uu}$	&	4158	&	0.79 & high-\Rey\\
26 341 & 23.6 &	349	&	698	& 2792	&	160	&	40	&	$0.3 \overline{uu}$	&	3329	&	0.84 & high-\Rey\\ 
  
\hline
29 048 & 30.7 &	111	&	222	& 890	&	60	&	15	&	$0.3 \overline{uu}$	&	4209	&	0.71 & high-\Rey\\
29 048 & 30.7 &	204	&	407	& 1628	&	105	&	26	&	$0.3 \overline{uu}$	&	3239	&	0.76 & high-\Rey\\
 \hline
  \end{tabular}
  \caption{Parameters of detections for each wall distances}
  \label{tab:3}
  \end{center}
\end{table}

\section{Universality of the convection velocity distribution in APG}
\begin{figure}
  \centerline{\includegraphics[width=1.0\textwidth]{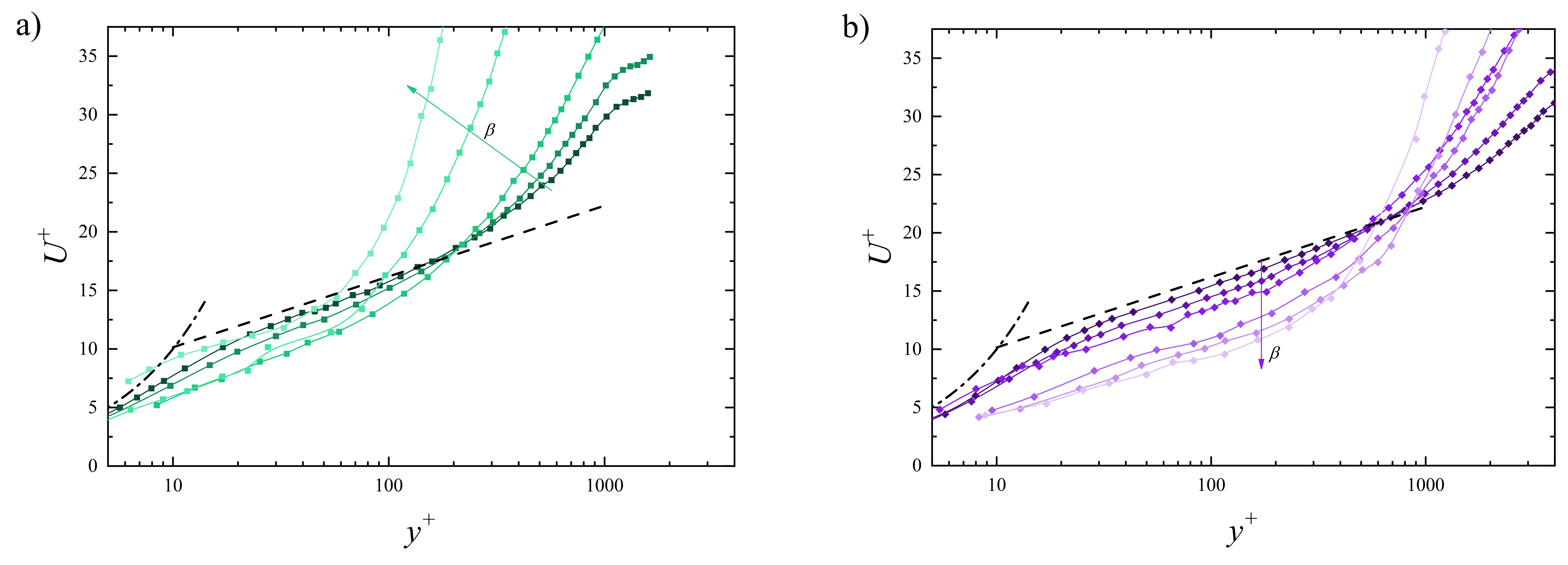}}
  \centerline{\includegraphics[width=0.5\textwidth]{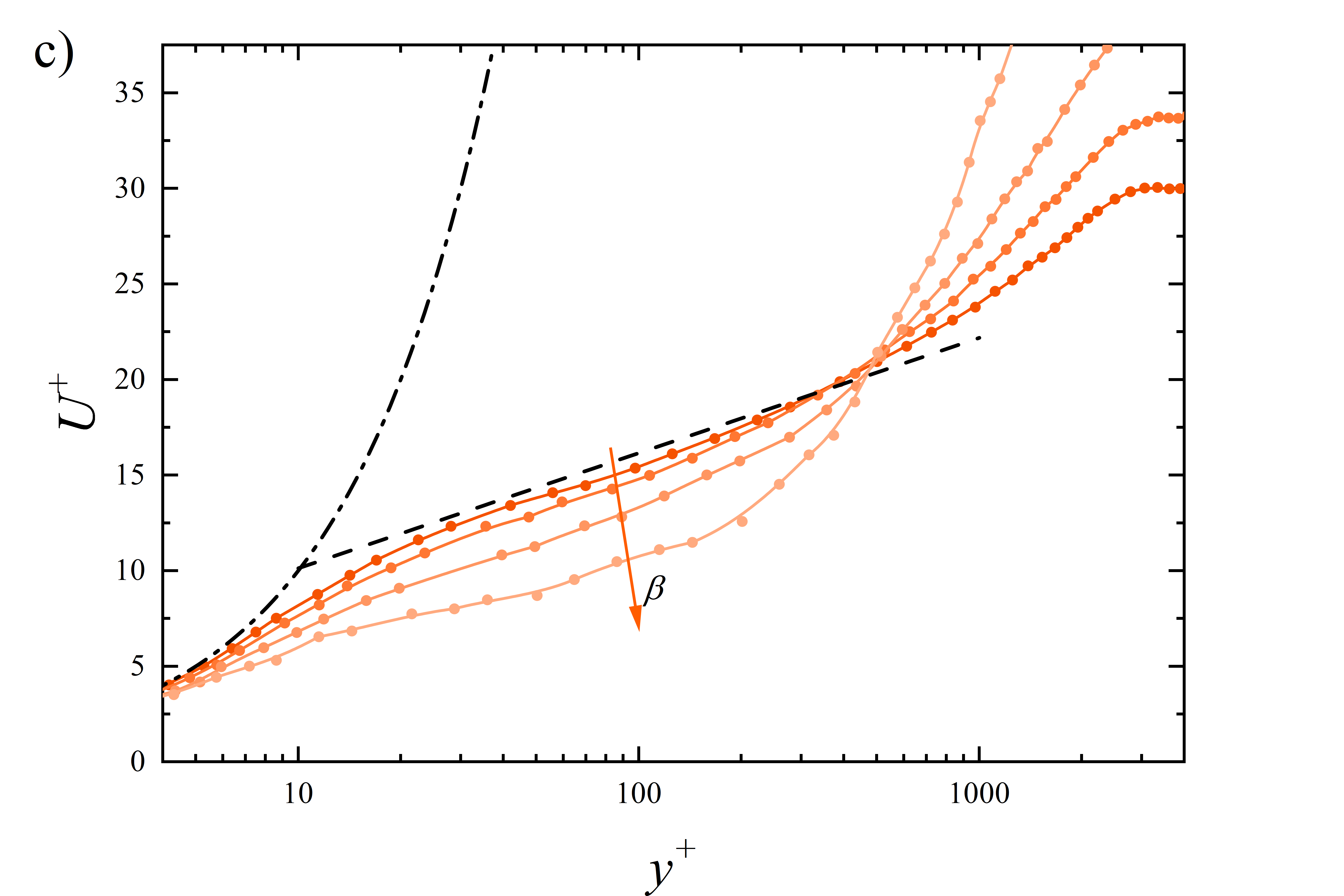}}
   \caption{Mean velocity profiles (in viscous units) in strong APG (a) low-$Re$ and (b) high-$Re$ and moderate APG (c). Symbols as in Table 1 and colour intensity decreases toward the separation.}
\label{fig:fig5}
\end{figure}

This section is intended to demonstrate the important role of the convection velocity in the description of the behaviour of TBL under APG conditions. The profiles of mean streamwise velocity component for $\beta \gtrsim 4$, plotted in classical inner scaling, are presented in Figure \ref{fig:fig5} for the low-$Re$ and high-$Re$ cases from Table \ref{tab:1}. Dashed line corresponds to logarithmic law (Eq. \ref{eq3}) and dashed-dot line corresponds to $U^+ = y^+$. The behaviour of the mean flow in APG, when scaled in wall units, is significantly different from the canonical ZPG flow. In the overlapping region, there is a universality of the mean velocity profile for the ZPG flows with the logline (\ref{eq3}), while in the APG there is a shift of the velocity profile below the logline, which is observed up to $\beta \approx 40$ \citep{Knopp2021}. After this value (close to the separation), the shift of the profile is diminishing, which is seen for highest $\beta$ profiles for the low-$Re$ case (see Figure \ref{fig:fig5}a). It appears that the effect is more pronounced with an increasing $Re$, which indicates that under APG the relative downstream decrease in friction velocity is weaker under the same external pressure gradient condition for both $Re$ cases \citep{Drozdz2021a}. This effect may be attributed to an increasing amplitude modulation at higher $Re$ flows (see Figure 9 in Ref. \cite{Drozdz2021a}). Moreover, as presented by \cite{Niegodajew2019}, there is a connection between LSM in the outer zone and the skin friction in APG flows. It was found that up to $\beta \approx 28$, the crossing between the mean profile and the logline occurs at the geometric centre of LSM, when $Re$ is sufficiently high. This indicates that the value of the mean velocity at the outer peak location of the turbulence intensity is proportional to $u_\tau$. This is an important observation as, despite  non-universality of the flow in the inner region (in the sublayer-buffer-overlapping layers), there is still a strong coupling between the sublayer and LSM from the overlapping layer. Such a  coupling between LSM and the skin friction indicates an influence of the amplitude modulation of the small scales and can be also found in other recent studies concerning ZPG flows \citep{Agostini2019} and APG flows \citep{yoon_hwang_sung_2018}, which additionally supports the finding of \cite{Niegodajew2019}. 

\begin{figure}
  \centerline{\includegraphics[width=0.7\textwidth]{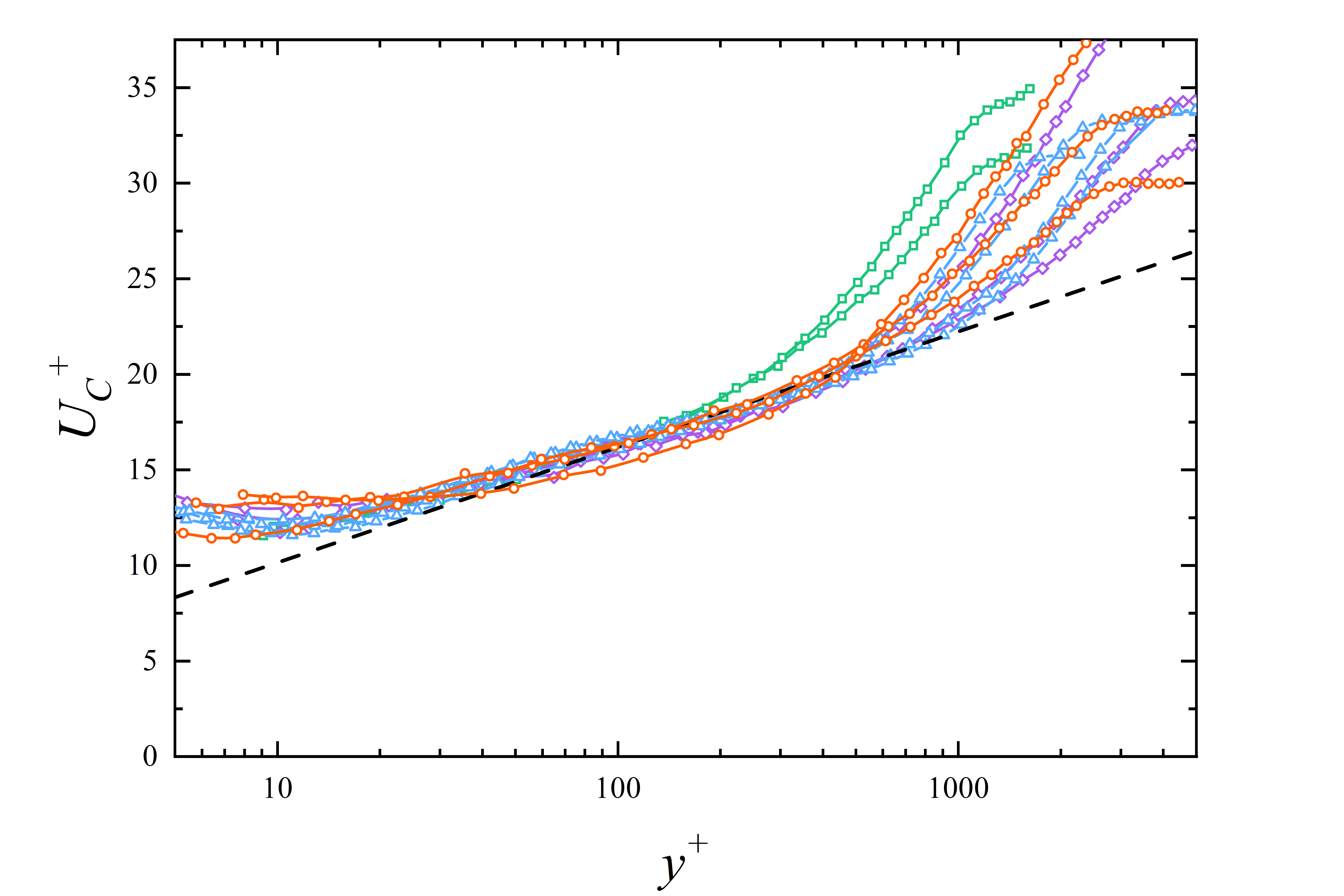}}
  \caption{Estimated convection velocity using relation \ref{eq2} with $\beta \lesssim 19$. Symbols as in Table \ref{tab:1}.}
\label{fig:fig6}
\end{figure}

Figure \ref{fig:fig6} presents the $U_C$ profiles scaled in the wall units for $4 \lesssim \beta \lesssim 19$. The dashed line corresponds to the logarithmic law (Eq. \ref{eq3}). It is shown that, despite an increased velocity deficit in APG, there is a universal behaviour of the $U_C^+$, which resembles the canonical logarithmic mean flow distribution in the wall units, even though it is not the case for the mean profiles (see Figure \ref{fig:fig5}). This also implies the proportionality of $u_\tau$ not to $U$ but to $U_C$ and is related to increased energy of sweep events, which is visible in Figures \ref{fig:fig3}b and \ref{fig:fig3}c. If there is an increased $u_\tau$ in the sublayer it should be related to the increased activity of small-scale vortices as the vorticity associated with small-scale vortices adds to produce the mean shear, however, it is not the case for APG flows. In this case the only explanation is the increased momentum transfer to the wall due to stronger small-scale sweeping, which to some extent compensates the declining vorticity that occurs near the wall under APG. It can be concluded that the decreased vorticity in APG gives the weaker mean shear, however, the sweeping strengthens the momentum transfer to the wall, so the universality of $U_C^+$ is maintained. 
In this sense, as was suggested by \citep{Drozdz2021a}, the LSMs affect the bursting process that enhance the small-scale sweep events and, thus, the mean friction velocity.

It should be noted that for the low-$Re$ case the logarithmic profile is barely visible or even does not occur. As it was show by in \cite{Drozdz2017Uc} for the $Re_\tau = 1000$ the $U_C^+$ profile in APG region is not universal. It seems that the universality of $U_C^+$ can be achieved only for higher Reynolds number TBLs, where the logarithmic region is observed on the mean velocity profile (here lowest $Re_\tau =1400$).

\begin{figure}
  \centerline{\includegraphics[width=0.7\textwidth]{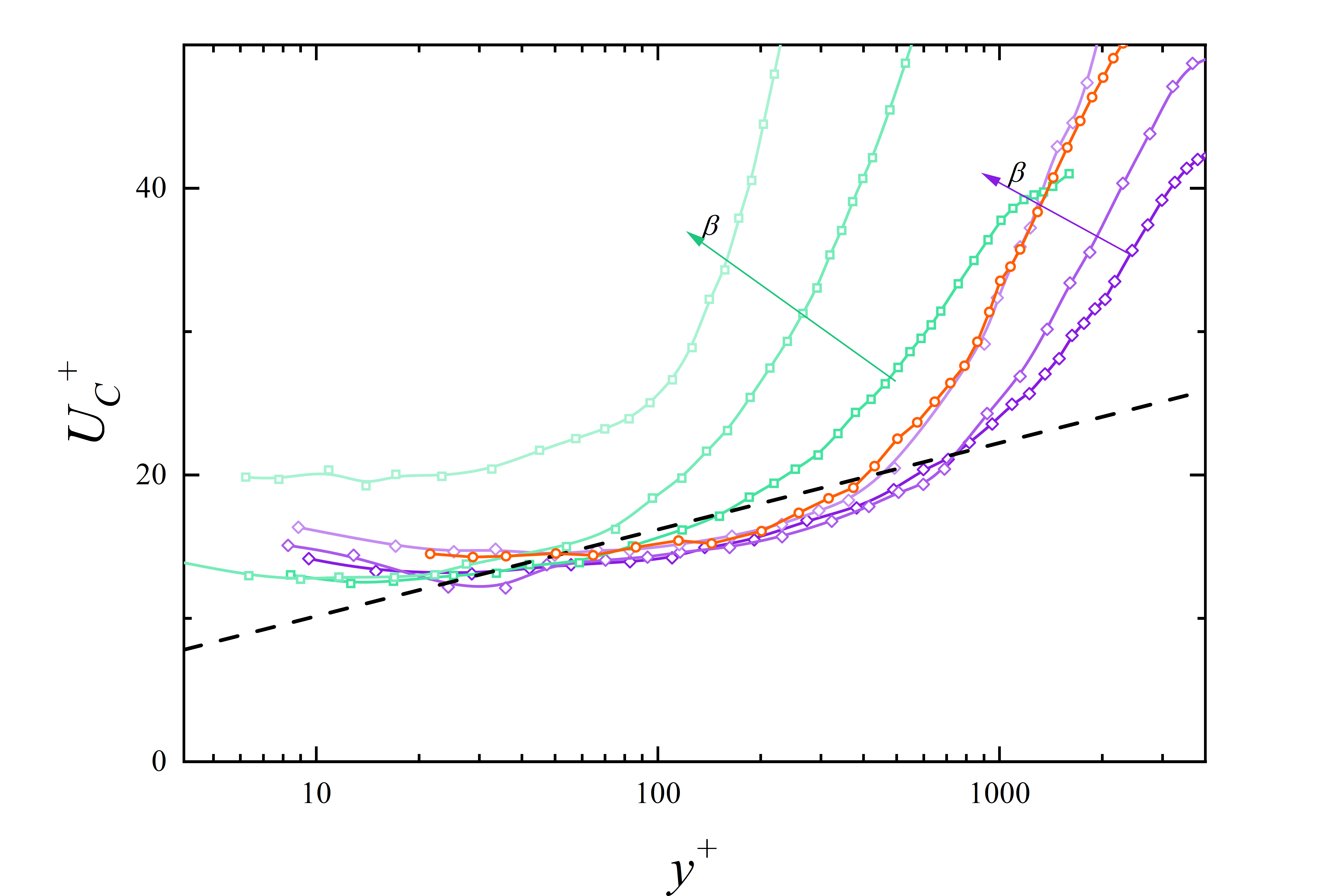}}
  \caption{Estimated convection velocity using relation \ref{eq2} with $\beta \gtrsim 19$. Symbols as in Table \ref{tab:1} and colour intensity decreases toward the separation.}
\label{fig:fig7}
\end{figure}

Figure \ref{fig:fig7} illustrates $U_C^+$ profiles obtained at $\beta \gtrsim 19$. One can see that the $U_C^+$ profiles can no longer fit to the universal logarithmic line, but they still surpass the mean velocity profiles, even to a greater extent than for the weaker APG (see the velocity profiles in Figure \ref{fig:fig5}), which indicates that the increased friction velocity is still observed. The compensation of the reduced small-scale vorticity by increased small-scale sweeping is still present but to a lesser extent close to separation.

The lower compensation of the decrease in skin friction due to mean and convection profile difference can be related to the extended backflow events that occur for high $\beta$ values. \cite{Vinuesa2017e} observed that, for $\beta \approx 14$, there is around 10\% of backflow and as previously suggested by \cite{Gungor2016}, the flow above such a $\beta$ value resembles the behaviour of a mixing layer in the upper half of large-velocity-defect boundary layers. Additionally, it was found that a mixing layer behaviour can be easily identified through the analysis introduced by \cite{Nagib2008}, in which the logarithmic-law parameters (namely the von Karman coefficient $\tilde{\kappa}$ and the intercept $\tilde{B}$, where $(\tilde{\cdot})$ indicates values obtained for each profile individually) were observed to follow the empirical relationship:

\begin{equation}
\tilde{\kappa}\tilde{B} = 1.6e^{0.1663\tilde{B}} - 1
\label{eq4}
\end{equation}
%

\begin{figure}
  \centerline{\includegraphics[width=0.7\textwidth]{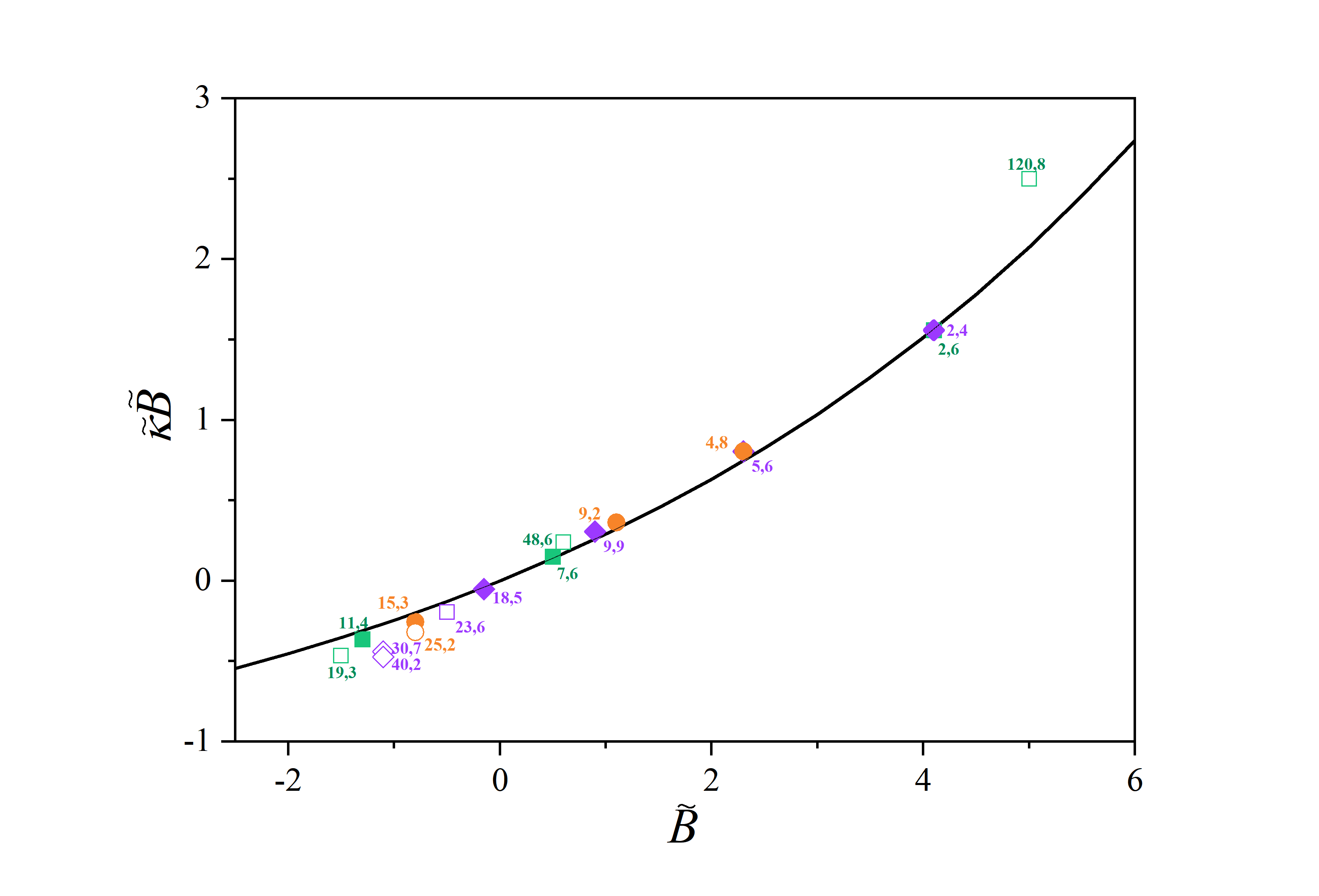}}
  \caption{Analysis of the $\tilde{\kappa}$ and $\tilde{B}$  log line parameters. Symbols as in Table 1 and the values of $\beta$ are shown in the label next to the symbols.}
\label{fig:fig8}
\end{figure}

The calculated values of $\tilde{\kappa}$ and $\tilde{B}$ for each $U^+$ profile from Figure \ref{fig:fig5} are shown in Figure \ref{fig:fig8} with $\beta$ values next to the symbols. It can be observed that this data closely follows the empirical relationship by \cite{Nagib2008} for all the cases characterised by $\beta \lesssim  19$. For higher $\beta$ values, the $\tilde{\kappa} \tilde{B}$ trend deviates from the line obtained using equation (\ref{eq4}). There is a cohesion between $U_C^+$ behaviour shown in Figures \ref{fig:fig6} and \ref{fig:fig7}, and the data shown in Figure \ref{fig:fig8}. Namely, the points characterised by $19 \lesssim \beta \lesssim 40$ are located below the empirical curve (Eq. \ref{eq4}) while the points relating to $\beta \gtrsim 40$ are positioned above this curve (see the open symbols in Figure \ref{fig:fig8}). It may be noted that the spread of the points with parameter $\tilde{B}$ is wider for the low-$Re$ case as the value of $\tilde{B}$ for the high-$Re$ case is not lower than ${\sim}0.5$, while for the low-$Re$ case the value of $\tilde{B}$ is at the level of $-1.5$ for profiles up to $\beta \approx 19$.

\section{Conceptual physical mechanism of the convection velocity impact on the mean flow}
According to \cite{DelAlamo2009}, in canonical flows, the $U_C$ profile can be approximated by the convolution of the mean velocity profile with a Gaussian window of the same size as a size of an eddy. This finding is based on the observation that the $U_C$ of an eddy is closely related to its dimension \citep{Townsend1976}. In canonical wall-bounded flows, $U_C$ is close to $U$ profile because vortices  near the wall are densely distributed. The term "vortices’/eddies’ density", herein $ED$, was introduced in Refs \citep{Wu2006b,Stanislas2008,Herpin2011,Herpin2013} and it refers to the number of eddies that occupy a given area. In the APG the approximation of \cite{DelAlamo2009} underestimates the mean convection velocity because it does not account for the lower activity of the small-scales located especially in low-speed zones. The mean convection velocity can be, however, approximated by the mean velocity that is calculated only in the areas that are densely occupied by the eddies. Such a conditional averaging procedure is not easy to preform in APG, since it requires a detection of the statistically representative vortical structures’ population for each wavenumber and for relatively constant pressure gradient conditions. Moreover, with the progress of APG, the areas with low energy dissipative vortices is growing, which cause that the $ED$ decreases. Within this process the mean turbulent profile, which is close to convection velocity profile at the beginning of APG, tends to the laminar one.

\begin{figure}
  \centerline{\includegraphics[width=1.0\textwidth]{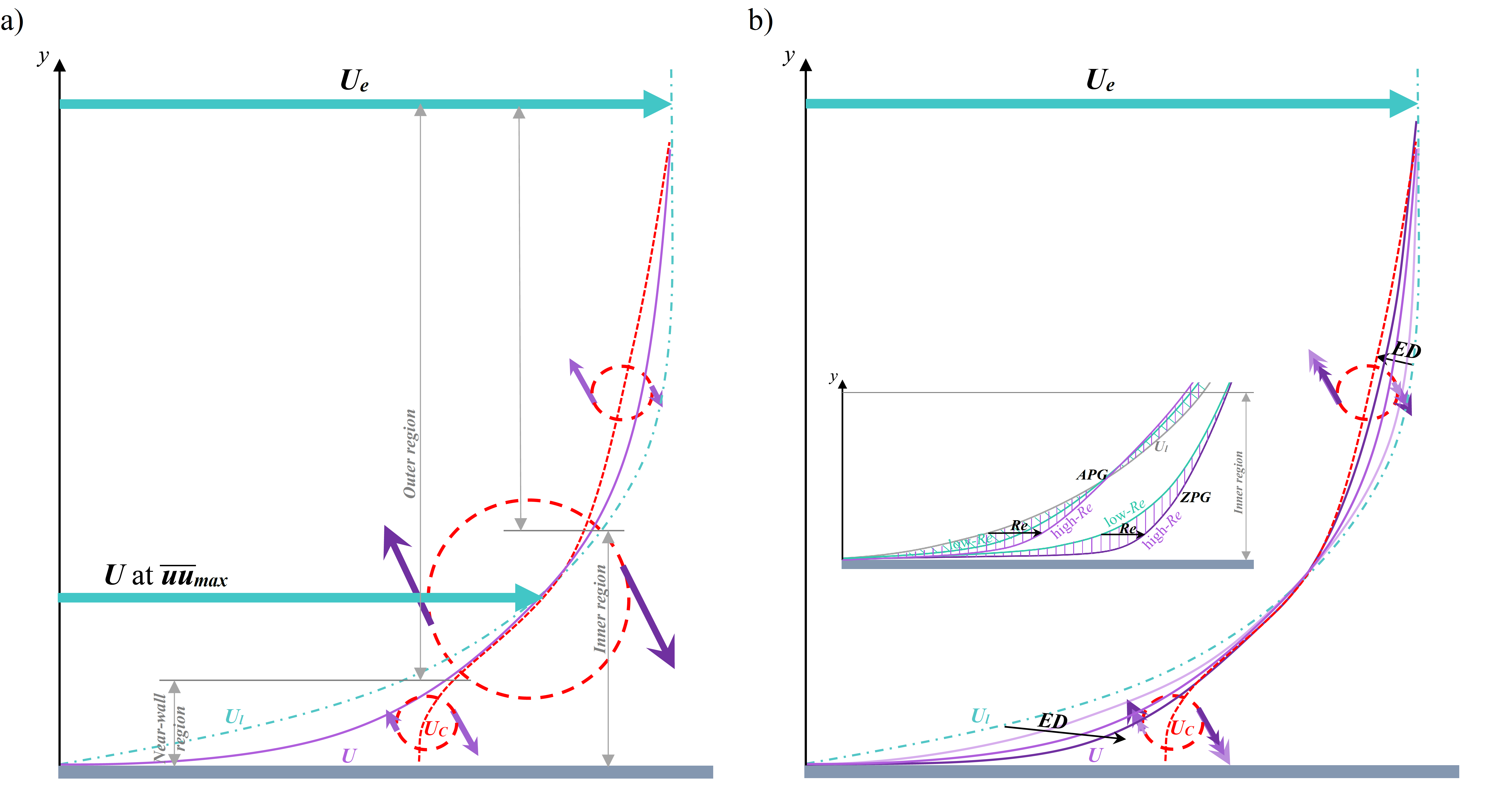}}
  \caption{Model of the convection velocity impact on the turbulent boundary layer profile (a), streamwise variation of the $ED$ in the APG with modification of momentum in the ZPG and APG with $Re$ due to the increase in $ED$ (b).}
\label{fig:fig9}
\end{figure}

The above discussion serves as a background for drawing a schematic representation of the turbulent mean profile that changes due to the $ED$. The general conceptual model of this process is illustrated in Figure \ref{fig:fig9}a. Here, the mean turbulent velocity profile is constructed as the weighted average of $U_C$ profile (which is conditionally averaged over the flow regions with a high $ED$) and pseudo-laminar profile $U_l$ (conditionally averaged over the flow regions with a low $ED$). The weighted average depends on the $ED$ difference between two flow regions (i.e. high- and low-momentum zones). Decrease in $ED$ is shown as the reduction in the colour intensity of the mean velocity profiles in Figure \ref{fig:fig9}b).

Generally, in APG (see Figure 9 in Ref. \citep{Drozdz2021a} shown in outer scaling), $U$ decreases as the mean shear in the overlapping region increases due to increased LSMs energy \citep{Drozdz2021a}. The conditionally averaged $U_l$ and $U_C$ also decrease due to increase in the LSMs energy. However, the decrease in $U_C$ is lower than that for $U$ \citep{Drozdz2021a}. Consequently, this results in an enhancement and weakening of the sweep and ejection events, respectively (see the changing intensity of the arrows that represent these events on the schematically presented near-wall vortex in Figure \ref{fig:fig9}b).

The enhancement of the sweeping in APG has already been documented in Refs \citep{Drozdz2017Uc,Gungor2016}, and it can also be observed on the phase-averaged waveforms in Figures \ref{fig:fig3}b and \ref{fig:fig3}c. 
The higher sweep, the higher momentum transfer to the wall and, thus, the higher the friction velocity. For weak and moderate pressure gradients (i.e. $\beta \lesssim 19$), the relative increase in the LSM energy is minor and, thus, the logarithmical universality of $U_C^+$ is observed (Figure \ref{fig:fig6}), which indicates that $u_\tau$ changes proportionally to $U_C$ (within the mentioned $\beta$ range). This proportionality is maintained by (as mentioned earlier) increased small-scale sweeping in the APG \citep{Drozdz2017Uc,Gungor2016} although the $ED$ is decreasing. For a strong pressure gradient, $U_C^+$ deviates from the logarithmic distribution, which is the result of the increase in LSM energy. Although the $ED$ decreases further, the friction velocity is still enhanced (as the $U_C^+$ profile in Figure \ref{fig:fig7} is located below the log-line) because of stronger sweeping. The effect of increased sweeping disappears at $\beta \approx 100$ due to a low $ED$, as $U_C^+$ profile is laying above the log-line (see Figure \ref{fig:fig7}) and $U$ profile is closer to log-line (see last profile in Figure \ref{fig:fig5}). 

As mentioned earlier, the stronger shift of the mean velocity profile below the log-line, which is visible for the high $Re$ case in Figure \ref{fig:fig5}, suggests that the small-scale sweeping increases with a growing \Rey, what is schematically demonstrated in Figure \ref{fig:fig9}b for both ZPG and APG. As can be seen, in the ZPG, the mean velocity profile becomes more filled towards higher \Rey. While in APG, this effect is visible in the near-wall region only. However, in the wake region, increased velocity deficit occurs with a growing \Rey, which is the effect of more filled $U$ profile near the wall. It is also observed that in this region the sweep event is dumped while the ejection is enhanced (see also Ref. \citep{Drozdz2011d}), however, these events have a low energy and they do not contribute much to the mean flow. The increasing \Rey, in this case, causes that the contribution of the sweep and ejection in the Reynolds shear stress tends to be the same. In the overlapping region, there is no effect either on the small-scale events or on LSM, this is because the overall convection velocity is equal to the mean one as it was concluded by \cite{Chung2010}.

\begin{figure}
  \centerline{\includegraphics[width=0.9\textwidth]{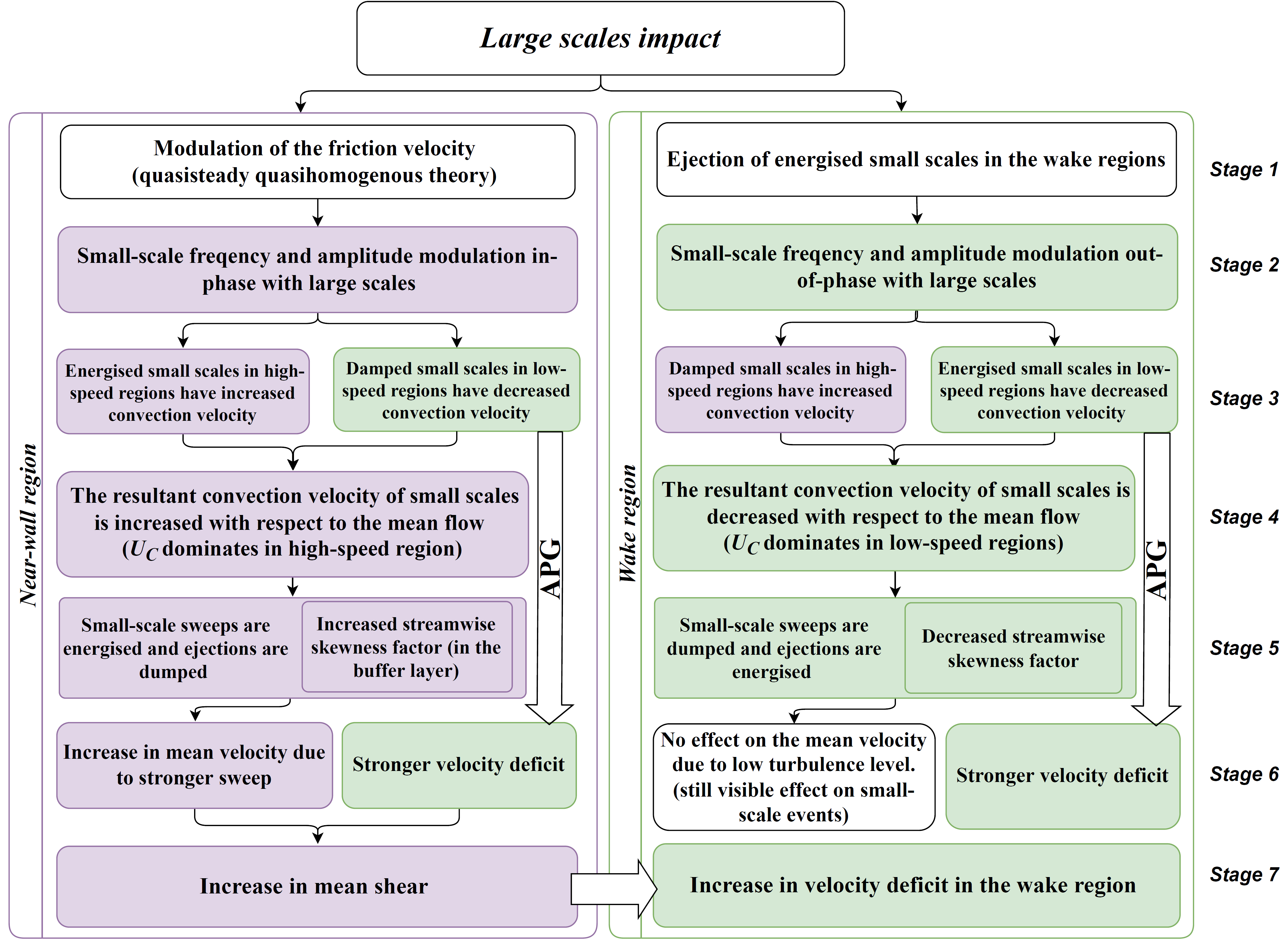}}
  \caption{Diagram of the small scales modulation that lead to a modification of the turbulent mean profile.}
\label{fig:fig10}
\end{figure}

Based on the above discussion, a conceptual diagram that describes the mechanism responsible for the modification of the mean turbulent profile due to LSMs is presented in Figure \ref{fig:fig10}. The diagram concerns the modification of the mean profile in both, the near-wall and wake regions within seven stages. The whole process is triggered by the LSMs that modulate the friction velocity in the near-wall region, according to the quasisteady quasihomogenous hypothesis of \cite{Zhang2016}, while in the outer part LSMs cause an ejection of the small-scales into the wake region (Stage 1). Consequently, the amplitude and frequency modulation occurs, which is in phase near the wall and out of phase in the wake region with LSM (Stege 2). This in turn induces modulation of small-scale $U_C$ with respect to the large-scale signal (Stage 3). The mean convection velocity (averaged over high- and low-speed regions) is enhanced near the wall, and it is reduced in the outer zone (Stage 4). This is because there are much more small-scale structures in the high-speed regions close to the wall, and thus, $U_C$ in that region contributes more to the mean $U_C$; this is why $U_C$ surpass the mean velocity. The opposite effect occurs in the wake region, in which the resultant convection velocity is lower than the mean velocity since the small scales are ejected into the wake region. The higher mean convection velocity, near the wall, enhances the sweep and dumps the ejection, which is also manifested by an increased skewness factor (Stage 5). The opposite effect on the sweep and ejection events occurs in the wake region. An enhanced sweep causes an increased momentum transfer to the wall and, hence, an increased friction velocity (Stage 6). In the wake region, there is no effect on the mean flow due to a low energy of the small-scale events, although the modulation still occurs. The final outcome, due to this process, is the increased mean shear in the TBL near the wall. Additionally, in APG, the mean velocity is reduced in the outer region (Stage 7). The process becomes stronger as $ED$ grows, which is due to the increased \Rey. In this sense, as was suggested by \citep{Drozdz2021a}, LSMs affect the bursting process by enhancing the small-scale sweep events and thus the mean friction velocity.

\section{Effect of the convection velocity on the mean flow near the separation}
\begin{figure}
  \centerline{\includegraphics[width=0.6\textwidth]{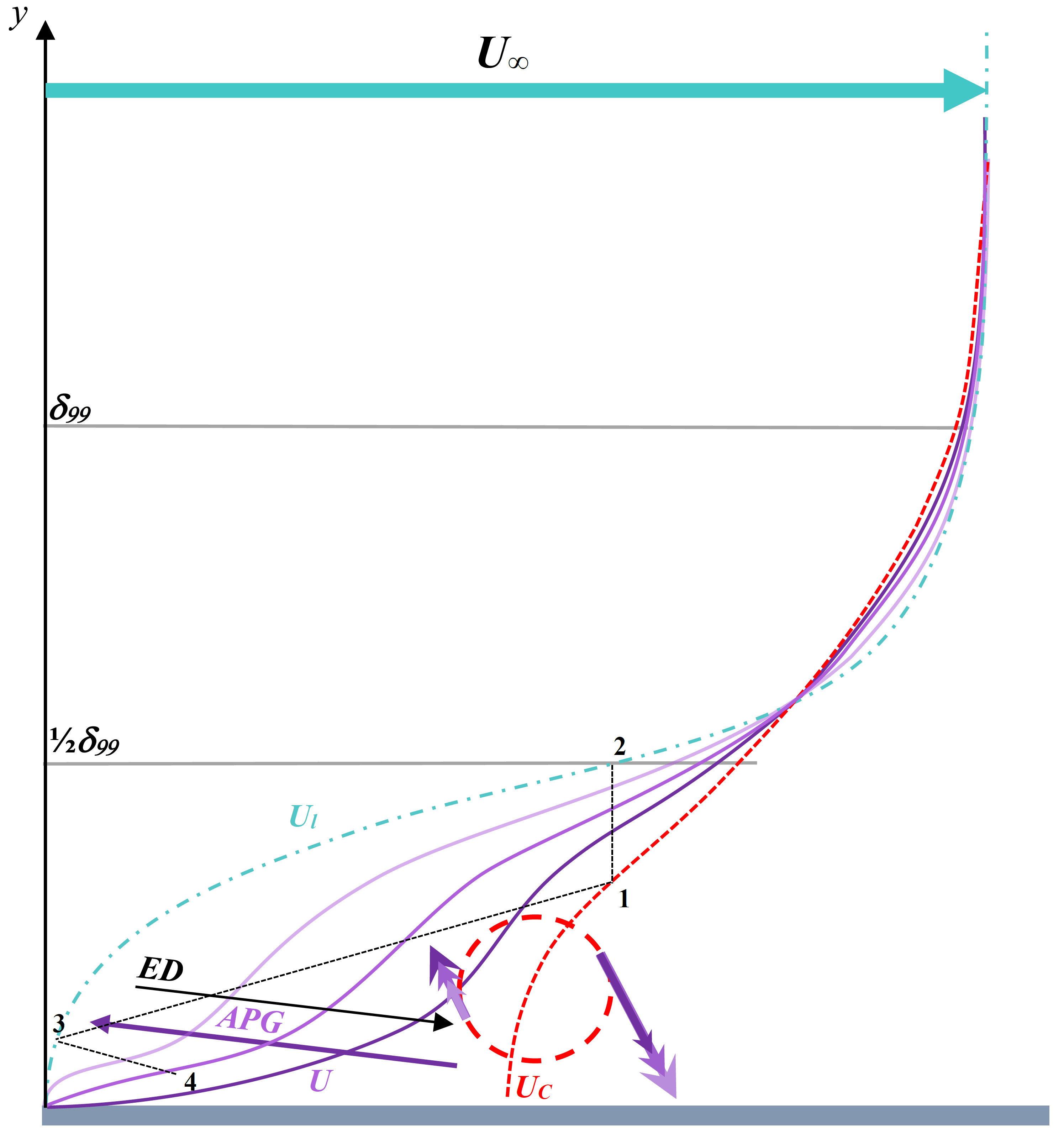}}
  \caption{Schematic representation of the convection velocity that constructs the turbulent boundary layer profile in the vicinity of the separation. Lines between the points from 1 to 4 indicate changes in the inflection points location.}
\label{fig:fig11}
\end{figure}

From a practical point-of-view, it is interesting to determine to what extent the mechanism responsible for increased convection velocity affects the flow close to the separation, where the inflection point in the central part of the mean velocity profile occurs. The occurrence of the inflection points close to the separation has already been discussed by \cite{Song2000}, \cite{Schatzman2017a} and \cite{Maciel2017}. The authors, however, did not provide a clear physical explanation on why the multiple inflection points appear on the mean profile. Plausible explanation is the increased convection that has a strong impact on the momentum transfer to the wall. The following schematic graphical interpretation illustrating the inflection points appearance on the mean profile is presented in Figure \ref{fig:fig11}. The interpretation is based on analysis of TBLs mean profiles at the vicinity of the separation, where there is a stronger mean shear in the overlapping region and even greater difference between the convection and mean profiles (see Figure 9 in \cite{Drozdz2021a}).

According to \cite{Drozdz2017Re,Drozdz2021a} at the visinity of separation the concave part of the mean profile can be clearly observed due to the presence of the inner inflection point at the location $y/\delta_{99} \approx 0.05$ and the outer one at the outer peak location of the maximum of streamwise Reynolds stress profile $\overline{uu}_{max}$ (discussed in more detail later in this section). \cite{Buckles1984} showed that, when the flow separates, the vortices sweep fluid toward the wall and it entrains the fluid from the reversed-flow region in the upward direction (increase in the momentum). This is correlated with large, positive skewness values in the reversed-flow region, which is caused by the passage of shear-layer vortices overhead. These observations led \cite{Buckles1984} to suggest that the detached shear flow is driven by a mechanism other than just the external pressure gradient. 
Also \cite{Maciel2006a} claimed “pressure force and the turbulent transport no longer play an important dynamic role close to the separation... [and the outer zone] ...becomes essentially as inertial flow”. 
The analysis supported by literature study proves that an increased sweep was already observed in previous works, however, the explanation of the physical mechanism of how it influence the flow has not been proposed.
According to the presented model the inflection point appearance is due to the increased convection of the turbulence, which induces an increased sweep (shown in Figure \ref{fig:fig11} with different colour intensities) and due to a stronger, compared to the ZPG, momentum transfer towards the wall at small wavelengths \citep{Drozdz2017Re,Drozdz2021a}. 

If $U_C$ were not higher than the $U$ close to the wall, only a single inflection point would be visible. However, in fact additional convex part appears which generates two inflection points. According to \cite{Gungor2016}, another inflection point can be visible in the vicinity of the wall and its presence becomes more visible with an increasing percentage of the backflow. This inflection point, however, is usually located so close to the wall that it cannot be measured experimentally, which was also the case in the present experiment.

In Figure \ref{fig:fig11} one can also observe how the changes of the inflection points location (dotted line) occur with respect to $ED$. Particularly, the concave part of the $U$ profile originates from the inflection point observed on the $U_C$ profile (point 1). The outer inflection point moves from point 1 to point 2, when the flow approaches the separation, which coincides with the RMS maximum of the fluctuation in the middle of the TBL thickness \citep{Gungor2016,Drozdz2017Re,Drozdz2021a}. The second dotted line (between points 1 and 3) indicates that the inner inflection point is modified with respect to $ED$. There is also a third dotted line (from point 4 to 3) that represents the additional inflection point close to the wall \citep{Gungor2016}, which should be visible within the vicinity of the separation. Below this line, there is a second concave part of the profile. On the 3rd point location the convex part of the profile disappears when convection no longer plays an important dynamic role. To the best of our knowledge the above explanation can be regarded as the first, so far, consistent physical description of multiple inflection points emergence. 

\section{Conclusions and discussions}
The paper examines the mean convection velocity of turbulent structures in the TBL exposed to weak, moderate and strong APG flows for a wide range of $Re_\tau \approx 1400-4000$. The convection velocities were estimated using an approach that employs the cross-product term $\overline{3\overline{u_S^2 u_L}}$ of the decomposed streamwise skewness factor validated with the free-of-wake-effect two-point correlation method. The obtained results showed that both approaches provide the comparable $U_C$ values, which confirms that the skewness-based method can be effectively used in APG flows. The results obtained in the present study confirm the previous findings of \cite{Drozdz2017Uc,Drozdz2017tp} and \cite{Balantrapu2021} concerning substantial increase in $U_C$  with respect to the mean flow.

The key finding of this work was that, in the overlapping region of APG flows, the convection velocity profiles (when scaled in viscous units) reassemble the universal logarithmic law characteristic for the ZPG flows up to $\beta \lesssim 19$ for the considered range of $Re$. An important conclusion that can be drawn from this observation is that $u_\tau$ in APG is not proportional to $U$ (as in ZPG) but to $U_C$ in the inner region of TBL. This effect, however, is not observed for the $U_C^+$ profiles obtained for higher $\beta$ values. Namely, for $19 \lesssim \beta \lesssim 40$, the $U_C^+$ profiles are shifted below the log-law line whereas for $\beta \gtrsim 40$ (close to the separation) a shift in the $U_C^+$ profiles above the log-law line is observed. This effect is also seen in the $\tilde{\kappa}\tilde{B}$ versus $\tilde{B}$ diagnostic diagram proposed by \cite{Nagib2008}. Namely, the $\tilde{\kappa}\tilde{B}$ values deviate from the empirycal line of \cite{Nagib2008} in an identical manner as the $U_C^+$ profiles from the universal log-line.

In general, towards stronger APG the difference between $U_C$ and $U$ increases which causes stronger sweeping that enhances momentum transfer to the wall and compensates the weaker mean shear profile that is created by lower vorticity near the wall in APG. That leads to the increase in friction velocity (except the near separation region) visible as the shift of the logarithmic mean profile $U^+$ below universal log-line in APG flows.

The analysis of our own data, supplemented by findings from the literature, provided the basis for formulation of the physical mechanism that explains the impact of increased convection velocity on the mean flow (due to induced eddies' density $ED$). According to this model, LSMs modulate $u_\tau$ in the near-wall region, according to the quasisteady quasihomogenous hypothesis of \cite{Zhang2016}. Consequently, the amplitude and frequency modulations occur, which are in phase with the LSM. This in turn induces $U_C$ modulation with respect to the large-scale signal and so the resultant $U_C$ (averaged over high- and low-speed regions) is enhanced (due to denser population of vortical structure in high speed regions). A higher mean $U_C$ enhances the sweep and dumps the ejection, which is also manifested by an increased skewness factor. An enhanced sweep causes an increased momentum transfer to the wall and, finally, an increased mean shear in TBL.  The process becomes more pronounced as $ED$ grows, so with increasing \Rey. It is also explained that the canonical TBL exhibits more momentum near the wall compared to the laminar one (due to variation in small-scale eddies’ density). The discussion, however, is not limited to the small velocity deficit in APG, but also to strong velocity deficit close to the separation where the velocity profiles exhibit multiple inflection points appearing close to turbulent separation.

As the universality of the convection velocity profiles is observed in the overlapping region, it can be concluded that the structure of the turbulent flow remains the same as for the ZPG, but increasingly more low-momentum zones become present towards stronger APG, which influence the mean velocity. The effect of the large and small scales modulation has to also be of great importance in this process, since an increase in the energy of the large-scale motion is responsible for the low- and high-momentum zones that are created. In APG flows, the turbulent structures are produced mostly in the high-momentum zones and, therefore, they have increased convection velocity. However, the mean velocity decreases due to the emergence of the severe low-momentum zones. 

It is expected that the proposed mechanism will shed new light on the paradigm of the turbulent mixing process in the non-canonical wall-bounded flows. Namely, an increase in the convection velocity, with respect to the mean one, enhances the sweep event in the inner region of TBL. Future research may concern a capturing of phase lag between LSM and envelopes of frequency, amplitude and convection velocity in APG as the proposed model does not address this issue.

\bigskip

\textbf{Acknowledgments}

We are grateful to W.J. Baars who provided Matlab script for linear coherence spectra calculations.

\bigskip

\textbf{Funding}

The investigation was supported by National Science Centre under Grant No. UMO-2020/39/B/ST8/01449

\bigskip
\textbf{Declaration of Interests} 

The authors report no conflict of interest.

\bibliography{convection}

\bibliographystyle{jfm}


\end{document}